%% file: 16-02-28-Jsac_CR.tex
\documentclass[10pt,twocolumn,twoside]{IEEEtran}
\IEEEoverridecommandlockouts

\usepackage[active]{srcltx} 
\usepackage{graphicx,color}
\usepackage{epsfig}
\usepackage{balance}
\usepackage{cite}
\usepackage{amssymb}
\usepackage[cmex10]{amsmath}
\usepackage{amsfonts}
\usepackage{amsthm}
\usepackage[applemac]{inputenc} 
\usepackage{subfigure} 
\usepackage{algorithm,algorithmic}

\input{macros}

\newtheorem{theorem}{Theorem}%[section]
\newtheorem{example}{Example}%[section]

\theoremstyle{remark}
\theoremstyle{definition}
\newtheorem{defin}{Definition}

\newtheorem{rem}{Remark}

\RequirePackage[normalem]{ulem} %DIF PREAMBLE
\RequirePackage{color}\definecolor{RED}{rgb}{1,0,0}\definecolor{BLUE}{rgb}{0,0,1} %DIF PREAMBLE
\begin{document}

\title{Speeding up Future Video Distribution via Channel-Aware Caching-Aided Coded Multicast }
\author{Angela Sara Cacciapuoti,~\IEEEmembership{Senior Member,~IEEE}, Marcello Caleffi,~\IEEEmembership{Senior Member,~IEEE}, Mingyue Ji,~\IEEEmembership{Member,~IEEE}, Jaime Llorca,~\IEEEmembership{Member,~IEEE} and Antonia Maria Tulino,~\IEEEmembership{Fellow,~IEEE}
\thanks{A. S. Cacciapuoti, M. Caleffi, and A. M. Tulino are with the DIETI Department, University of Naples Federico II, Italy. E-mail: \{angelasara.cacciapuoti, marcello.caleffi, antoniamaria.tulino\}@unina.it}
\thanks{M. Ji is with Broadcom Limited, CA. Email:  \{mingyue.ji@broadcom.com\}}
\thanks{J. Llorca and A. M Tulino are with Bell Labs, Nokia, NJ. Email:  \{jaime.llorca, a.tulino\}@nokia.com}
\thanks{This work was supported in part by the PON projects  ``FERSAT: studio di un sistema di segnalamento FERroviario basato sull'innovativo utilizzo delle tecnologie SATellitari e della loro integrazione con le tecnologie terrestri" and DATABANC  ``CHIS: Cultural Heritage Information System", and in part by the Campania POR project  ``myOpenGov". }}

\maketitle

%-----------------------------------------------------------------------------------------------------------------------------------------------------------------------------
% ABSTRACT
%-----------------------------------------------------------------------------------------------------------------------------------------------------------------------------
\begin{abstract}
Future Internet usage will be dominated by the consumption of a rich variety of online multimedia services accessed from an exponentially growing number of multimedia capable mobile devices. 
As such, future Internet designs will be challenged to provide solutions that can deliver bandwidth-intensive, delay-sensitive, on-demand video-based services over increasingly crowded, bandwidth-limited wireless access networks. 
One of the main reasons for the bandwidth stress facing wireless network operators 
is the difficulty to exploit the multicast nature of the wireless medium when wireless users or access points rarely experience the same channel conditions or access the same content at the same time. In this paper, we present and analyze a novel wireless video delivery paradigm based on the combined use of channel-aware caching and coded multicasting that allows simultaneously serving multiple cache-enabled receivers that may be requesting different content and experiencing different channel conditions. 
To this end, we reformulate the caching-aided coded multicast problem as a joint source-channel coding problem 
and design an achievable scheme that preserves the cache-enabled multiplicative throughput gains of the error-free scenario,   
%while guaranteeing independent maximum per-receiver (access point) channel rates, unaffected by the presence of receivers with worse channel conditions. 
{by guaranteeing per-receiver rates unaffected by the presence of receivers with worse channel conditions}. 
\end{abstract}

%-----------------------------------------------------------------------------------------------------------------------------------------------------------------------------

%-----------------------------------------------------------------------------------------------------------------------------------------------------------------------------
% INTRODUCTION
%-----------------------------------------------------------------------------------------------------------------------------------------------------------------------------
\section{Introduction}
\label{sec:1}
The latest projections \cite{Cisco1,Cisco2} suggest that, by 2019, mobile data traffic will increase nearly tenfold with respect to 2014, accounting for nearly two-thirds of the total data traffic. Furthermore, it is predicted that nearly three-fourths of the mobile data traffic will be video by 2019. In line with these trends, this paper considers the design and analysis of a novel wireless video delivery paradigm that specifically addresses two of the major predicted shifts of the Future Internet, i.e., \textit{from fixed to mobile} and \textit{from cable video consumption to IP video consumption}, by pushing caching to the wireless edge and exploiting the multicast nature of the wireless medium via channel-aware coded multicasting.

Recent works have addressed the design of heterogeneous wireless access networks composed of a combination of macro-, micro-, and pico-cells integrated within the cellular infrastructure, with the main advantage of the increased spatial reuse resulting from the simultaneous localized high-bandwidth wireless connections from small cell base stations to user devices. However, in many cases, the high cost incurred in wiring the small cells results in a shift of the bandwidth bottleneck to the wireless backhaul that serves the multiple access points. To alleviate this problem, recent works have proposed the use of a caching directly at the wireless edge, e.g., at wireless access points or end user devices, with the goal of reducing both latency and wireless backhaul requirements when serving video content (see \cite{GolMolDimCai-13} and references therein). 
In this context, recent information theoretic studies have shown that the use of network coding over the wireless backhaul can significantly improve the performance of wireless caching networks by creating cache-enabled coded multicast transmissions useful for multiple users (access points) even if requesting different content.
In fact, is has been shown that the use of wireless edge caching and coded multicasting enables multiplicative caching gains, in the sense that the per-user throughput scales linearly with the local cache size \cite{MadNie12,maddah2013decentralized,ji2014average, ji2015random,Pr:JiTulLloCai14,ji2015Efficient,ji2015Preserving,shanmugam2014}.
However, the underlying assumption in existing information theoretic literature on caching networks is that the channels between the content source and the users either exhibit the same qualities or follow a shared error-free deterministic model. In practice, wireless channels are affected by impairments, such as multipath and shadow fading, and they must be modeled as non-deterministic noisy channels. Furthermore, different wireless channels can exhibit significant differences in their qualities, i.e., admissible throughputs, even when the users are closely located. This channel heterogeneity can significantly degrade the performance of any multicast-based approach in which the worst-channel user dictates the overall system performance  \cite{Chou03,ScottJSTSP15}.

\begin{figure}[t]
\centering
\includegraphics[width=0.85\linewidth]{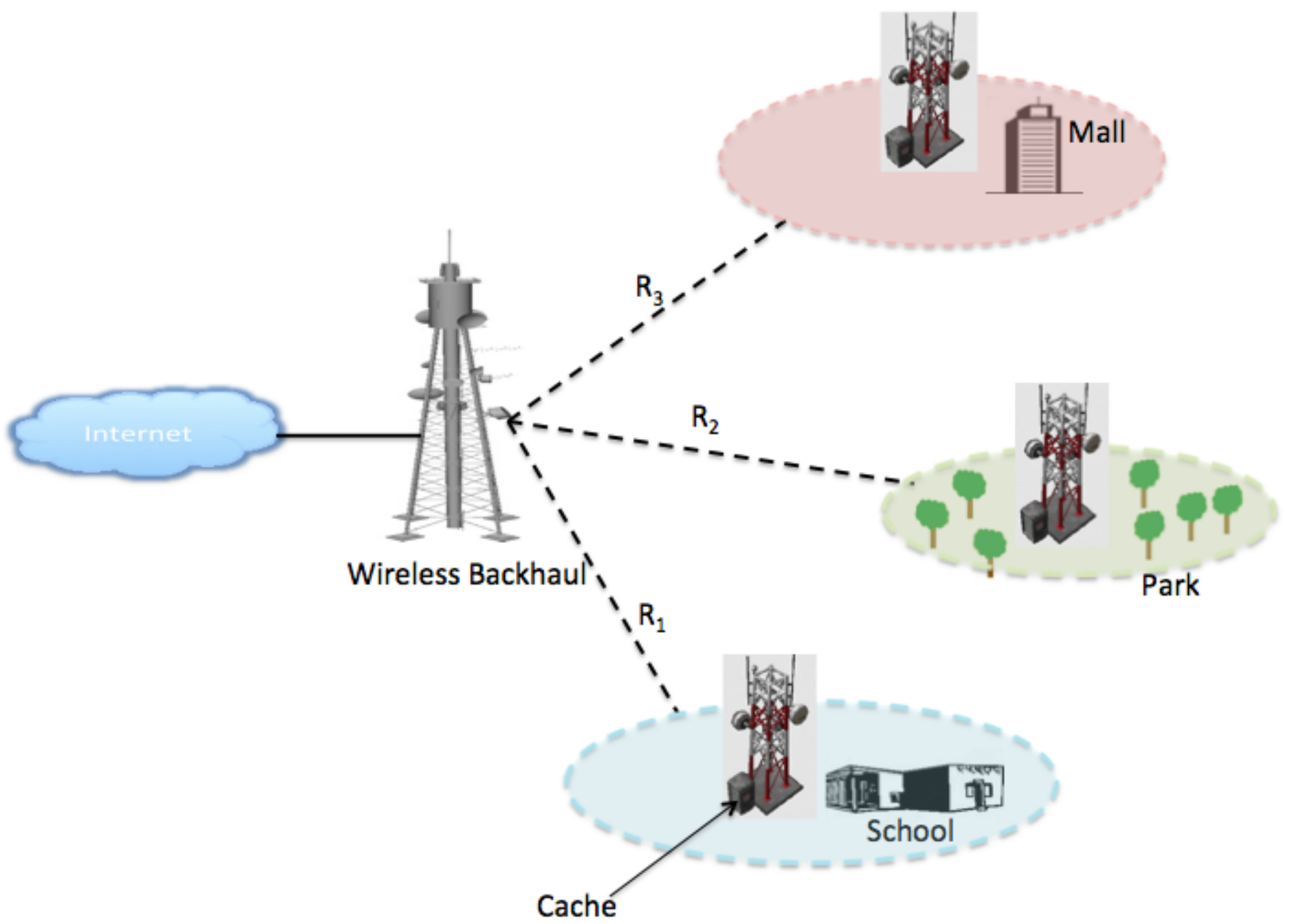}
\caption{Network Model}
	\label{fig_model}
\end{figure}

In the following, we address these open problems by considering a heterogeneous wireless edge caching architecture, as shown in Figure~\ref{fig_model}. It is composed of a macro-cell base station (BS) that distributes video content to a number of mobile devices, with the help of dedicated cache-enabled small cell base stations or wireless access points, referred to as \textit{helpers}. Specifically, by exploiting the helper caching capabilities, the BS distributes the video content through coded multicast transmissions, reducing both content distribution latencies as well as overall network load \cite{GolMolDimCai-13,Pr:JiTulLloCai14}. We model the channel between macro BS and helpers as a stochastically degraded broadcast channel.  
Our approach is based on formulating the caching-aided coded multicast problem over a noisy broadcast channel as a \textit{joint source-channel coding problem},  
in order to optimize the sharing of bandwidth and caching resources among cache-enabled receivers downstream of a common broadcast link with heterogeneous channel conditions.  
Our contributions can be summarized as follows:
\begin{itemize}
	\item An information theoretic framework for wireless video distribution that takes into account the specific characteristics of the wireless propagation channel in the presence of any combination of unicast/multicast transmission and wireless edge caching;
	\item A channel-aware caching-aided coded multicast video delivery scheme, referred to as Random Popularity based caching and Channel-Aware Chromatic-number Index Coding (RAP-CA-CIC), that guarantees 
	{a rate to each receiver within a constant factor of the optimal rate had the remaining users experienced its same channel conditions,} 
	%the highest  admissible video throughput to each receiver (helper) for the given propagation conditions, 
	i.e., completely avoiding throughput penalizations from the presence of receivers with worse propagation conditions. 
	\item A polynomial-time approximation of RAP-CA-CIC, referred to as Random Popularity based caching and Channel-Aware Hierarchical greedy Coloring (RAP-CA-HgC) with running time at most cubic in the number of receivers and quadratic in the number of (per-receiver) requested video descriptions.
\end{itemize}

The rest of the paper is organized as follows. Section \ref{sec:model} describes the network and video distribution models. Section~\ref{sec:4} presents an information-theoretic formulation for the caching-aided wireless video delivery problem over a degraded broadcast channel. In Sections~\ref{sec: Caching Placement Scheme} and ~\ref{sec: Multicast Scheme}, we specialize the aforementioned information-theoretic formulation for a specific choice of  the cache encoder (Sec. \ref{sec: Caching Placement Scheme}) and of the joint source-channel multicast encoder (Sec. \ref{sec: Multicast Scheme}), introducing the proposed RAP-CA-CIC and RAP-CA-HgC algorithms. Detailed descriptions of CA-CIC and its polynomial-time approximation CA-HgC are given in Section \ref{algs}. In Section \ref{sec: specialdec}, we describe the multicast decoder and the overall achievable performance. 
Section~\ref{sec:6} presents simulation results for the proposed and state of the art schemes, along with related discussions. We conclude the paper in Section~\ref{sec:7}.

%-----------------------------------------------------------------------------------------------------------------------------------------------------------------------------
% Section 2
%-----------------------------------------------------------------------------------------------------------------------------------------------------------------------------

\section{System Model}
\label{sec:model}

\subsection{Network Model}
\label{sec:2}

We consider a wireless broadcast caching network consisting of a sender node (base station) and $U$ receiver nodes (helpers) $\mathcal{U}=\{1,\ldots, U\}$,  
as shown in Figure \ref{fig_model}. The sender has access to a content library $\mathcal{F}=\{1,\ldots,m\}$ containing $m$ files, where each file has entropy $F$ bits.  
Each receiver $u \in \Uc$ has a cache with storage capacity $M_u F$ bits (i.e., $M_u$ files). 
We denote by $M_{u,f}$ the fraction of file $f \in \mathcal{F} $  stored at receiver $u$, such that $\sum_f M_{u,f} \leq M_u$. 
Without loss of generality, the files are represented by binary vectors $W_f \in \FF_2^F$. 

Differently from existing works for this network \cite{MadNie12,maddah2013decentralized,ji2014average, ji2015random, Pr:JiTulLloCai14,ji2015Efficient,ji2015Preserving}, here the sender is connected to the receivers via lossy links. Specifically, we model the links between the sender and the receivers as stochastically degraded 
binary broadcast channel (BC).  The processes governing  the links time evolution are  assumed to be {stationary and ergodic}.  
While the binary field is specially convenient for ease of presentation, our approach can be extended to  general stochastically degraded broadcast channels \cite{Gam79}, with the case of arbitrary additive noise broadcast channels being particularly immediate. Our analysis includes both a binary symmetric broadcast channel  (BS-BC) and a binary erasure  broadcast channel (BE-BC).
\begin{itemize}
\item{\textbf{BS-BC Case}}: The channel output $Y_u[t]$ observed by the $u$-th receiver at the $t$-th channel use takes values in the binary alphabet  $\mathcal Y \equiv \{0,1\}$ and is given by
\begin{equation}
Y_u[t]= X[t]+ Z_u[t],
\end{equation}
where $X[t]$ denotes the binary encoded symbol sent by the sender at the $t$-th channel use, and $Z_u[t]$ denotes the additive noise of the channel corresponding to receiver $u$, modeled as a Bernoullian variable with parameter given by the channel degradation $\epsilon_u$, i.e., $Z[t] \sim \mathcal B (\epsilon_u)$. 
Letting $\eta_u$ denote the achievable channel rate, we have 
\begin{equation}
\eta_u \leq  1- H(\epsilon_u),
\label{Ac_rate_BS}
\end{equation} 
with $H(\epsilon_u)$ denoting the binary entropy function.
 \item{\textbf{BE-BC Case}}: The channel output $Y_u[t]$ observed by the $u$-th receiver at the $t$-th channel use exactly reproduces the channel input $X[t]$ with probability $(1- \epsilon_u)$ and otherwise indicates an erasure event, with probability $\epsilon_u$. 
In this case, $Y_u[t]$ takes values in the ternary alphabet $\mathcal Y \equiv \{0,1, \star \}$ so that an erasure event is represented by the erasure symbol "$\star$". The achievable channel rate $\eta_u$ then satisfies
\begin{equation}
\eta_u \leq  1- \epsilon_u.
\label{Ac_rate_BE}
\end{equation}
\end{itemize}

%\DIFdelbegin \DIFdel {Without loss of generality, we assume that $0\leq \epsilon_1 \leq \epsilon_2 \leq \ldots \leq \epsilon_U \leq 1$.}\DIFdelend

\subsection{Video Distribution Model}
\label{sec:3}

We consider a video streaming application, in which each file $f\in\mathcal F$ represents a video segment, which is multiple description coded\footnote{As will be shown later, compared to scalable video coding \cite{SchMarWie07}, multiple description coding offers significant advantages that are especially relevant for the use of caching-aided coded multicasting.} into $D$ descriptions using, for example, one of the coding schemes described in  \cite{Soljanin11,Chou03}. Each description is packaged into one information unit or packet for transmission. Each packet is represented by a binary vector of length (entropy) $B=F/D$ bits. A low-quality version of the content can be reconstructed once a receiver is able to recover any description. The reconstruction quality improves by recovering additional descriptions and depends solely on the number of descriptions recovered, irrespective of the particular recovered collection. Hence, there are $D$ video qualities per segment, where the entropy in bits of quality $d\in\{1,\dots, D\}$, containing $d$ descriptions, is given by $F_d = B\,d$. 
%\red{We assume time is slotted with slot duration $\Delta$. Without loss of generality, the system is designed such that a single user with LFU caching can receive $F bist in $\Delta time units
{Note that in video streaming applications, every video segment has the same (playback) duration, which we denote by $\Delta$ in time units (or channel uses). Hence, the difference in quality levels solely depends on the number of bits $F_d$.  
In practice, the video segment duration should be chosen according to the maximum quality offered by the video application and the channel conditions \cite{videobuf}. 
In this paper, we set $\Delta = \gamma F$, with $\gamma> 0$ being a system parameter that depends on the channel conditions.}

Video segments are characterized by a popularity or \textit{demand distribution} $\Qm = [q_{f,u}], u=1, \cdots, { U}, f=1, \cdots, m$, where $q_{f,u} \in [0,1]$ and $\sum_{f=1}^m q_{f,u} = 1$ (e.g., receiver $u$ requests file $f$ with probability $q_{f,u}$). 
%Without loss of generality up to index reordering, we assume $q_{f,u}$ has non-increasing components $q_{1,u} \geq \cdots \geq q_{m,u}$. 
Let  $\fsf_{u}$  denote the random request at receiver $u$. The realization of $\fsf_{u}$ is denoted by $f_{u}$. 

It is important to note that this paper considers a general video on-demand setting, in which receiver requests follow an arbitrary popularity distribution. As such, the demand message set cannot be represented as a \emph{degraded message set} \cite{KorMar77,Gal74} since a given receiver's demand is not necessarily a subset of another receiver's demand.  
Our demand model includes and generalizes any combination of degraded message sets, via possible overlapping of receiver demands, and message cognition, via available cached or side information \cite{Mansour}. 

We consider a video delivery system that operates in two phases: a caching phase\footnote{Because of the time-scale separation between caching and transmission phases, the caching phase is sometimes referred to as the placement or pre-fetching phase.} 
 followed by a transmission phase.
\begin{itemize}
\item{\textbf{Caching Phase:}} 
The caching phase occurs during a period of low network traffic. In this phase, using the knowledge of the demand distribution and the cache capacity constraints, the sender decides what content to store at each receiver. 
{We denote by $\mu_{u,f}$ the number of descriptions of file $f$ cached at receiver $u$.}
\item{\textbf{Transmission Phase:}} 
After the caching phase, the network is repeatedly used in a time slotted fashion with time slot duration $\Delta$,   
%\red{set to the the time (in channel uses) it takes for the best channel receiver to decode $F$ bits (highest quality video segment).}\footnote{\red{In video streaming applications, this is typically referred to as the \emph{buffer time} or \emph{playback delay} \cite{}}.}
{given by the video segment duration.} % of the buffer at each receiver set by the video streaming application to absorb the channel fluctuation. }
%given by the playback duration of the buffer at each receiver set by the video streaming application to absorb the channel fluctuation. }
Each receiver requests one video segment per time slot. 
Based on the receiver requests, the stored contents at the receiver caches, and the channel conditions, the sender decides at what quality level to send the requested files (i.e., how many descriptions to send to each receiver), encodes the chosen video segments into a codeword, and sends it over the broadcast channel to the receivers, such that each receiver can decode its requested segment (at the scheduled quality level) within $\Delta$ {channel uses}. We denote by $d_u$ the scheduled quality level for receiver $u$. 
\end{itemize}

%-----------------------------------------------------------------------------------------------------------------------------------------------------------------------------
% Section 3
%-----------------------------------------------------------------------------------------------------------------------------------------------------------------------------

\section {Problem Formulation}
\label{sec:4}
As in previous studies for the broadcast caching network, the goal is to characterize the rate-memory region defined as  the closure of the set of all achievable rate-memory tuples. 
While a number of studies have characterized this fundamental tradeoff under error-free channel conditions  \cite{MadNie12,maddah2013decentralized,ji2014average, ji2015random,Pr:JiTulLloCai14,ji2015Efficient,ji2015Preserving,shanmugam2014}, only a few recent works have provided first steps towards the characterization of the rate-memory region under a degraded broadcast channel \cite{AsaOngJoh15, TimWig15}. These studies are however limited to scenarios with only two receivers and unrealistic constraints such as requiring the worst channel receiver to have larger storage capacity. In the following, we introduce an information-theoretic formulation for the general wireless video delivery over a stochastically degraded broadcast channel problem described in Section \ref{sec:model}.

\subsection {Information-Theoretic Formulation}
As stated earlier, we represent the video segments by binary vectors $W_f \in \FF_2^F$ of  entropy $F$. At the beginning of time, a realization of the library $\{W_f \}$\footnote{For ease of exposition, in the following, unless specified, we denote by $\{A_i\}$ the full set of elements $\{A_i : i\in\mathcal I\}$.} is revealed to the sender. A $ \left( \Delta,  \{\epsilon_u\} \right)$-delivery scheme consists of:

\subsubsection{Cache Encoder}
The sender, {given the knowledge of the demand distribution $\Qm$ and the cache sizes $\{M_u\}$}, fills the caches of the $U$ receivers through a set of $U$ encoding functions $\{Z_u: \FF_2^{mF} \rightarrow \FF_2^{{ M_u}F}\}$,  such that $Z_u(\{W_f\})$ denotes the codeword stored in the cache of receiver $u$.  

\subsubsection{Joint Source-Channel Multicast Encoder}
%Given the time slot duration $\Delta$,  
The sender, given the knowledge of the cache configuration $\{ Z_u\}$, the channel conditions  $\{\epsilon_u\}$, {the network time slot duration $\Delta$}, and the receiver requests $\fv$, schedules the quality level $d_{u}$ for each requested video segment  through a multicast encoder, defined by a variable-to-fixed encoding function
$ {\bf X} : \FF_2^{mF} \times \Fc^{{U}} \rightarrow \FF_2^{\Delta}$ (where $\FF_2^{\Delta}$ denotes the set of binary sequences of finite length $\Delta$)\footnote{The symbol $\FF_2^{\Delta}$ is used to indicate that the codeword length is fixed and dictated by the time slot duration $\Delta$.},  
such that the transmitted codeword is given by
\begin{equation}
\label{eq:encoder} {\bf X}\eqdef {\bf X} (\{Z_u\}, \{\epsilon_u\},  \mathbf{f}) , 
\end{equation} 
{where $ \mathbf{f}\eqdef [f_1, f_2, \ldots, f_U]$, with $f_u \in \mathcal{F}$,  
denotes the realization of the receiver random request vector ${\mathbf \fsf}= [\fsf_1, \fsf_2, \ldots, \fsf_U]$. }

\begin{rem} 
We remark that the classical separation source channel coding over compound channel (SSC-CC), while optimal for equal channel conditions, it is known to be suboptimal in general. 
\end{rem}

\begin{rem} 
%Stemming from the described Multicast Encoder, it results that 
Note that the variable-to-fixed nature of the multicast encoder is due to the fact that, while the length of the transmitted codeword  ${\bf X}$ is fixed to the network time slot duration $\Delta$, %given by the video segment playback duration, 
the amount of  information bits transmitted by the encoder is variable and  given by $\sum_u d_u B$. 
\end{rem}

\subsubsection{Multicast Decoders}
Each receiver $u\in\mathcal U$, after observing its channel output ${\bf Y}_u$, decodes the requested file $W_u \in \mathcal F$ as  $\widehat{W}_u = \lambda_u(\mathbf{Y}_u,Z_u,\mathbf{f})$, where  $\lambda_u : \mathcal{Y}^{\Delta} \times \FF_2^{M_uF} \times \Fc^{U }\rightarrow \FF_2^{F}$ denotes the decoding function of receiver $u$ (with $ \mathcal{Y}^{\Delta}$ denoting the set of the received sequences). 

The worst-case %\footnote{The $\sup_{\{W_f\}}$ indicates the worst-case over the file library.}
(over the file library) error probability of a  $\left( \Delta,  \{\epsilon_u\} \right)$-delivery scheme over a stochastically degraded broadcast  channel  with  channel degradations {$\{\epsilon_u\}$} % \DIFdelbegin \DIFdel { $0\leq \epsilon_1 \leq \epsilon_2 \leq \ldots \leq \epsilon_U \leq 1$}\DIFdelend 
 is given by 
\begin{align} 
\label{perr}
P_e^{(F)} ({\bf X}, \{Z_u\}, \{\lambda_u\}, \{\epsilon_u\}) =  \!\!
\sup_{\{W_f \}}   \PP \left ( \bigcup_{u \in \Uc} \Big \{ {\widehat{{W}}_u \neq  {W}_u } \Big \} \right ). 
\end{align}

\begin{defin}
\label{def:admissible}
A sequence of $\left( \Delta,  \{\epsilon_u\} \right)$-delivery schemes is called {\em admissible} 
%\footnote{\textcolor{red}{The condition established in Definition~\ref{def:admissible} is an  asymptotic condition. Clearly in such an asymptotic regime, if $F \rightarrow \infty$, then  also $\Delta \rightarrow \infty$.}} 
if  $$\lim_{F \rightarrow \infty} P_e^{(F)} ({\bf X}, \{Z_u\}, \{\lambda_u\}, \{\epsilon_u\}) = 0. $$ 
\hfill $\lozenge$
\end{defin}

\subsection{Average Performance Measures}

We define the average %\footnote{Throughout this paper, we directly use per-receiver delivery rate to refer to the average per-receiver delivery rate defined by \eqref{average-rate-user} and explicitly use average (or expected) per-receiver delivery rate if needed for clarity.}  
{\em per-receiver delivery rate}  of a $ \left( \Delta,  \{\epsilon_u\} \right)$-delivery scheme as the average number of bits per channel use provided to receiver $u$ via the combined use of the its own cache and the multicast codeword:\footnote{Note that \eqref{average-rate-user} indicates the worst-case (over the library) average (over the demands) \textit{per-receiver delivery rate}.}  
%of the segment delivered to receiver $u$. % within $\Delta$ time units. Specifically, the average {\em per-receiver delivery rate} is given by
\begin{equation} 
\label{average-rate-user} 
R_u ^{(F)} =  \inf_{\{W_f\}} \; \frac{B}{\Delta} \left( \E[d_u( \mathbb{\fsf}) ]  +\E[\mu_{u,\fsf_u}]\right),  
\end{equation}
where $d_u(\mathbb{\fsf})$ and $\mu_{u,\fsf_u}$\footnote{Recall that $\fsf_u$ denotes the random variable that models the request of receiver $u$.} are the number of descriptions delivered to receiver $u$ via the multicast codeword and via the receiver's cache, respectively, and the expectations are with respect to the random request vector $\mathbb{\fsf}$. 
%and finally  $t^*_u( \mathbb{\fsf})$ is the time needed to receiver $u$ to collect the $d_u(\mathbb{\fsf})$ descriptions} .
Note that  $d_u( \fsf) + \mu_{u,\fsf_u}\leq D$. 
%\textcolor{red}{Note that, in \eqref{average-rate-user}, $0 \leq t^*_u \leq \Delta$, since according to the video distribution model, receiver $u$ have to decode the requested segment (at the corresponding quality level) within $\Delta$ time units.} 
 
 %\textcolor{red}{The definition of average {\em per-receiver delivery rate} agrees with the intuition. In fact, when for example the storage capacity of the receiver cache is equal to the content library, $R_u ^{(F)}$ goes to infinity, as expected. In fact, $t^*$ in \eqref{average-rate-user} is equal to zero, since receiver $u$ has already all the descriptions of the file it requested. Hence there is no transmission at all. }
 
In addition, we can define the average {\em network delivery rate} of a $\left( \Delta,  \{\epsilon_u\} \right)$-delivery scheme as \begin{equation} 
\label{average-rate}
R^{(F)} = \inf_{\{W_f\}} \;  \frac{B }{\Delta}  \sum_{u=1}^U  \left(\E[d_u(\mathbb{\fsf}) ]  + \E[\mu_{u,\fsf_u}] \right).
\end{equation}

%\textcolor{red}{\begin{rem}
%At the denominator of  \eqref{average-rate} there is $\Delta$, since as described in Section~\ref{sec:3}, the network is repeatedly used in a time slotted fashion with time-slot duration $\Delta$ time units. Hence also if a receiver consumes a segment in less than $\Delta$ time units, a new round of requests starts after $\Delta$ time units.
%\end{rem}}

Similarly, we define the average  {\em per-receiver distortion} $\delta_u^{(F)}$  and 
the average  {\em network distortion} $\delta^{(F)}$ of a $\left( \Delta,  \{\epsilon_u\} \right)$-delivery scheme as  
\begin{equation} 
\label{average-dist-user}
\mathcal \delta_u^{(F)} = \sup_{\{W_f\}} \;  \left (1- \frac{E[d_u(\mathbb{\fsf})]  + \E[\mu_{u,\fsf_u}]  }{D}    \right), 
\end{equation}
 
\begin{equation}   
\label{average-dist}
\mathcal \delta^{(F)} = \sup_{\{W_f\}} \;  \left (1-  \frac{1}{U}\sum_{u=1}^U  \left(\frac{E[d_u(\mathbb{\fsf})] + \E[\mu_{u,\fsf_u}] }{D} \right)\right).
\end{equation}
 
\hfill

\begin{defin}
\label{defadmissible}
A set of average  {\em per-receiver distortions}   $(\delta_1,\ldots, \delta_U)$ is achievable if there exists a sequence of admissible 
$\left( \Delta,  \{\epsilon_u\} \right)$-delivery schemes with  
 average {\em per-receiver distortions} $(\delta_1 ^{(F)},\ldots, \delta_U ^{(F)})$ such that  
\begin{equation} 
\label{average-rateachivable}
\limsup_{F \rightarrow \infty } \delta_u^{(F)}\leq  \delta_u, \forall u \in \mathcal{U}. 
\end{equation}
\hfill $\lozenge$
\end{defin}

In the following, we focus on a particular class of admissible $\left( \Delta,  \{\epsilon_u\} \right)$-delivery schemes, based on a specific choice of  {\textbf{Cache Encoder}} and joint  source-channel {\textbf{Multicast Encoder}}. 
Specifically, we consider a cache encoder based on random fractional caching \cite{ji2015random} and a multicast encoder consisting of a channel-aware index coding scheme that builds on a novel graph coloring algorithm based on \emph{maximal generalized independent sets}. We refer to our solution as Random Popularity Caching and Channel-Aware Chromatic Index Coding (RAP-CA-CIC). RAP-CA-CIC is designed to assure that \emph{i)} all the receivers are able to decode the requested information from the received multicast codeword, 
and \emph{ii)} no receiver gets affected by receivers with worse channel conditions.

%-----------------------------------------------------------------------------------------------------------------------------------------------------------------------------
% Section 4
%-----------------------------------------------------------------------------------------------------------------------------------------------------------------------------

\section{Cache Encoder}
\label{sec: Caching Placement Scheme}

The cache encoder exploits the fact that a video segment is multiple description coded into $D$ descriptions, each of length $B=F/D$ bits. In the following, such descriptions are referred to as packets and denoted by $\{W_{f,\ell}: \ell=1, \ldots, D\}$.  
The caching phase works as follows.
Each receiver, instructed by the sender, randomly selects and stores in its cache a collection of $p_{f,u} M_u D$ distinct packets of file $f \in \mathcal{F}$, where ${\bf{p}}_u = (p_{1,u}, \ldots, p_{m,u}) $ is a vector  with components $0 \leq p_{f,u}\leq1/M_u, \forall f$, such that $\sum_{f=1}^m p_{f,u}=1$, referred to as the caching distribution of receiver $u$. Hence, the arbitrary element $p_{f,u}$ of ${\bf{p}}_u$ represents the fraction of the memory $M_u$ allocated to file $f$.
{The resulting codeword $Z_u(\{W_f\})$ stored in the cache of receiver $u$ is given by
\begin{align}
\nonumber
Z_u&=\left(W_{1,\ell_{1,1}^u}, \ldots, W_{1,\ell_{1,p_{1,u}M_u D}^u}, W_{2,\ell_{2,1}^u}, \ldots, W_{2,\ell_{1,p_{2,u}M_u D}^u},\right.\\ & \left. ,\ldots,  
W_{m,\ell_{m,1}^u}, \ldots,W_{m,\ell_{m,p_{m,u}M_u D}^u}\right) , 
\label{eq:cache}
\end{align}
where $\ell_{f,i}^u$ is the index of the $i$-th packet of file $f$ cached by receiver $u$. The tuples of indices $(\ell_{f,1}^u, \ldots, \ell_{f,p_{f,u}M_u D}^u)$ are chosen independently across for each receiver $u \in \mathcal U$ and file $f \in \mathcal F$, with uniform probability over all ${D \choose p_{f,u}M_u D }$ distinct subsets of size $p_{f,u}M_u D$ in the set of $D$ packets of file $f$. 
We refer to the random collection of cached packet indices across all receivers as the cache configuration, denoted by $\Msf=\{Z_u\}$. A given realization of $\Msf$ is denoted by $\Mm$. 
Moreover, we denote by $\Mm_{u,f}$ the vector of indices of the packets of file $f$ cached by receiver $u$. 

The aforementioned caching policy is referred to as \textit{RAndom fractional  Popularity-based (RAP) caching}, which is completely characterized by the caching distribution $\Pm=\{{\bf p}_{u}\}$. RAP is synthesized in Algorithm~\ref{alg1}.

\begin{algorithm}
\caption{RAndom Popularity-based (RAP) caching}
\label{alg1}
 {\small
\begin{algorithmic}[1]
\FORALL{$f \in \mathcal{F}$}
\STATE Each receiver $u\in\mathcal U$ caches $p_{f,u} M_u D$ distinct packets of file $f$ chosen uniformly at random. 
\ENDFOR
\RETURN $\Mm = [\Mm_{u,f}], u\in\{1, \dots, U\}, f\in\{1, \dots, m\}$.
\end{algorithmic}
}
\end{algorithm}}

\begin{figure}[t]
\centering
\includegraphics[width=1\linewidth]{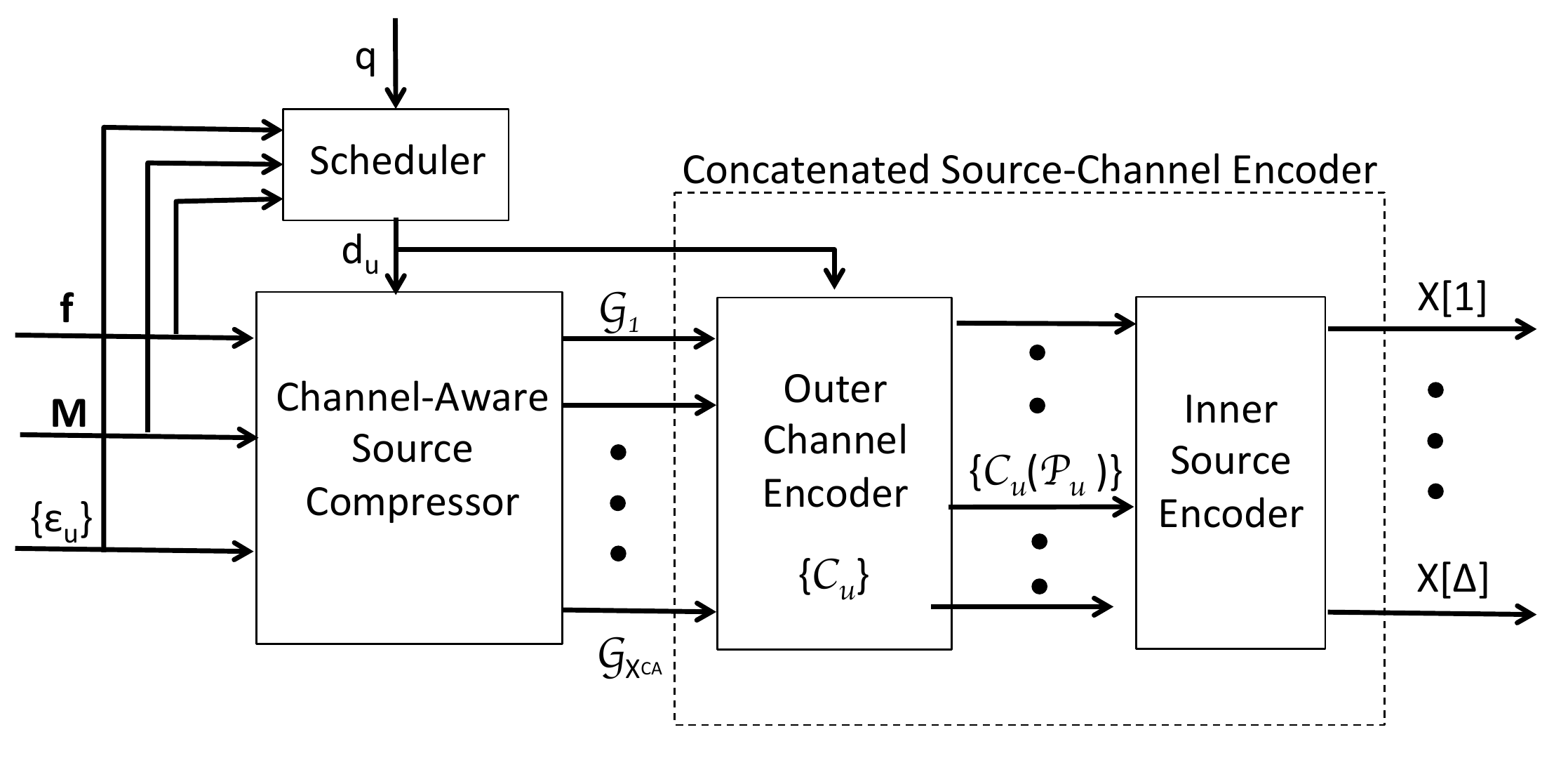}
\caption{Joint Source-Channel Multicast Encoder}
	\label{fig_encoder}
\end{figure}

We remark that a key property of RAP is that by choosing the identity of the packets to be cached uniformly at random, it increases the number of distinct packets cached across the network. The fact that receivers cache different packets of the same file is essential to enable coded multicast opportunities during the transmission phase \cite{ji2015random}. This is one of the reasons why multiple description coding is much more suitable than scalable video coding, since  randomly cached packets (descriptions) can always be used for decoding higher quality video versions. With scalable video coding, randomly cached descriptions may be wasted if not all previous descriptions are delivered during the transmission phase \cite{parisa}. 
While this can be avoided by only caching descriptions up to the maximum number that can be delivered to a given receiver, 
{the resulting reduced amount of overall cached information can lead to} % can creates too much overlapping in cached information, which in turn 
reduced coded multicast opportunities. In addition, any deviation due to imperfect knowledge of demand and channel conditions can again lead to wasted cached information, reducing system robustness. 

%-----------------------------------------------------------------------------------------------------------------------------------------------------------------------------
% Section 5
%-----------------------------------------------------------------------------------------------------------------------------------------------------------------------------

\section{Multicast Encoder }
\label{sec: Multicast Scheme}

The goal of the joint source-channel multicast encoder is to create a multicast codeword that allows each receiver to decode its requested video segment at a  rate that is equal to the maximum rate achievable if the remaining $U-1$ receivers had its same channel degradation.
The proposed multicast encoder, depicted in Fig.~\ref{fig_encoder}, is composed of  three main building blocks: %$i)$ a \emph{scheduler}, $ii)$ a {\em channel-aware source compressor}, and $iii)$ a {\em concatenated source-channel encoder}.

\subsection{Scheduler}
\label{descriptions}

Recall that, for a given realization of the packet-level cache configuration $\Mm$ and the channel rates $\{\eta_u\}$, the  joint source-channel multicast encoder is a variable-to-fixed encoder that, at each realization of the request vector $\fv$, encodes the scheduled descriptions for each receiver $\{d_u\}$ into a multicast codeword $\Xm$ of fixed length $\Delta$.
The scheduler in Fig. \ref{fig_encoder} computes the number of descriptions scheduled for receiver $u$ as
\begin{equation}
\label{eq:descriptions}
d_u(\fv) = \frac {\Delta \, \eta_u}{B \, \psi(\fv,\Mm)},
\end{equation}
where  $ \psi(\fv,\Mm)$ is a function of the  packet-level cache realization $\Mm$ and of the request vector realization $\fv$, whose expression is given by Eq.~\eqref{eq: chromatic number 2} in Appendix~\ref{sec: Proof of Theorem up}. 

As will be shown in Appendix \ref{sec: Proof of Theorem up}, Eq.~\eqref{eq:descriptions} guarantees that, with a multicast codeword of length  $\Delta$, receiver $u$ obtains $d_u$ descriptions (each of size $B$) at a rate $\eta_u/ \psi(\fv,\Mm)$, which  is the maximum (within a constant factor)  rate achievable when the remaining $U-1$ receivers have the same channel degradation $\eta_u$ \cite{ji2015random}.  

Note from \eqref{eq:descriptions} that the number of descriptions depends on $\psi(\fv,\Mm)$ and consequently on the cache and demand realizations. 
{This dependence can be specially critical for %As shown in Fig. \ref{fig_encoder}, 
the concatenated source-channel encoder, % takes $\{d_u\}$ as input, hence the dependence of $d_u$ on the demand realization can be specially critical, 
possibly requiring channel codebook updates at each request round (see Section \ref {subsec:concatenated}). }
In order to eliminate this overhead, we impose that receiver $u$ obtains $d_u$ descriptions at a rate $\eta_u/ \psi(\fv,\Mm)$, but set the codeword length given by  $d_u B \psi(\fv,\Mm) /\eta_u$ to be, \emph{in average}, equal to the network time slot duration $\Delta$. In this case, the number of scheduled descriptions for receiver $u$ is given by
\begin{equation}
\label{eq:descriptionsAv}
d_u =  \frac {\Delta \, \eta_u}{B \, \E[\psi(\mathbb{\fsf},\Msf)]},
\end{equation}
where $\E[\psi(\mathbb{\fsf},\Msf)]$ is the expectation of $\psi(\mathbb{\fsf},\Msf)$ taken over the random request vector $\fsf$ and random cache configuration $\Msf$, whose expression is given by Eq. \eqref{eq:chi} in Appendix \ref{sec: Proof of Theorem up}. 

\begin{rem}
Note that, although not explicitly stated in Eqs. \eqref{eq:descriptions} and \eqref{eq:descriptionsAv}, the number of scheduled descriptions for receiver $u$ must be bounded by $D(1- M_{u,f_u})$.\footnote{Recall that $f_u$ denotes the realization of the file requested by receiver $u$.}  
In addition, the ratios {$\frac {\Delta \eta_u}{B \psi(\fv,\Mm)}$} and $\frac {\Delta \, \eta_u}{B \, \E[\psi(\mathbb{\fsf},\Msf)]}$ may not be integer and, in practice, would need to be rounded down. 
\end{rem}

\subsection{Channel-Aware Source Compressor} 

The goal of the channel-aware source compressor is to cluster the set of packets (descriptions) scheduled for each receiver $\{d_u\}$ into a smaller set of equivalent classes. 
A key concept driving this process is what we refer to as {\em generalized independent set} (GIS). As shall be clear from its formal definition in Section \ref{algs}, the concept of GIS generalizes the classical notion of independent set in the graph coloring literature. The proposed compressor clusters the entire set of scheduled packets (descriptions) into the minimum number of GISs satisfying the following conditions:
\begin{itemize}
\item
Any two packets in a GIS scheduled for different receivers can be transmitted in the same time-frequency slot without affecting decodability.
\item
The set of packets in a GIS scheduled for receiver $u$, denoted by $\mathcal P_u$, satisfies
\begin{equation}
\label{eq:2_1}
\frac{|\mathcal P_u|}{\eta_u}= \frac{|\mathcal P_i|}{\eta_i}, \quad \forall u,i \in \{1, \ldots, U\}. 
\end{equation}
Note that \eqref{eq:2_1} makes sure that each receiver is scheduled a number of packets (descriptions) proportional to its channel rate.
\end{itemize}

As described in detail in Section~\ref{sec:CAC}, this minimization corresponds to a NP-hard optimization problem related to finding the minimum number of GISs that cover a properly constructed \emph{conflict graph}. We refer to the minimum number of GISs needed to cover the conflict graph as \emph{channel-aware chromatic-number}, $\chi_{\text{CA}}$,  
and to the associated transmission scheme as {\em Channel-Aware Chromatic-number Index Coding (CA-CIC)} (see Section \ref{sec:CAC}). Given the exponential complexity of CA-CIC, we then provide in Section~\ref{sec:hgc} a practical polynomial-time approximation of CA-CIC, referred to as {\em Channel-Aware Hierarchical greedy Coloring (CA-HgC)}.

\subsection{Concatenated Source-Channel Encoder} 
\label{subsec:concatenated}
The concatenated source-channel encoder generates the multicast codeword $\bf X$ in which the scheduled descriptions $\{d_u\}$  are encoded. It consists of an outer channel encoder and an inner source encoder. The outer channel encoder generates channel codewords of length $n$, while the inner source encoder generates the final multicast codeword $\bf X$ of length $\Delta$, by concatenating linear combinations of $n$-length channel  codewords.

\subsubsection{Outer channel encoder} 
As illustrated in Fig.~\ref{fig_encoder}, the outer channel encoder takes as input the GISs generated by the channel-aware source compressor
and, for each GIS and receiver $u$, encodes the descriptions associated with the set $\mathcal P_u$ of entropy $|\mathcal P_u| B$ bits, into a codeword $\mathfrak{C}_u(\mathcal P_u)$ of length $n$, via channel codebook $\mathfrak{C}_u$, generated according to the following definition: 
\begin{defin}
\label{def_codebook}
An $(2^{n \eta_u}, n )$ code for the $u$-th binary channel consists of the following:
\begin{itemize}
\item An index set $\{1,2, \ldots, 2^{n  \eta_u}\}$;
\item An encoding function 
\begin{equation}
\label{eq:coding}
\mathfrak{C}_u: \{1,2, \ldots, 2^{n  \eta_u}\} \rightarrow  \{0,1\}^n 
\end{equation}
yielding the codebook $\{\mathfrak{C}_u(1), \ldots \mathfrak{C}_u(2^{n  \eta_u})\}$;
\item A decoding function
\begin{equation}
\label{eq:decoding}
g_u:   \mathcal{Y}^n  \rightarrow \{1,2, \ldots, 2^{n  \eta_u}\};
\end{equation}
\end{itemize}
\hfill $\lozenge$
\end{defin}

\begin{rem}
Note that if $\frac{|\mathcal P_u|}{\eta_u} B$  is not a multiple of $n$, then zero-padding is applied to the $|\mathcal P_u| B$ bits.
\end{rem}

The sender notifies the computed codebooks to each receiver at network setup or anytime channel conditions change. Hence, each receiver is aware, not only of its own codebook, but also of the codebooks of the other receivers. 

In line with Section \ref{descriptions}, and as shown in Appendix \ref{sec: Proof of Theorem up},  in order to guarantee that each receiver obtains $d_u$ descriptions at a rate $\frac {\eta_u}{\psi(\fv,\Mm)}$, the channel codeword length is set to $n= \frac {\Delta}{\psi(\fv,\Mm)}$ (see Eq. \eqref{eq:descriptions}). The fact that $n$ depends on the demand realization implies that the channel codebook may need to be updated and notified to each receiver at each request round, leading to significant overhead. 
In order to eliminate this overhead, we set $n= \frac {\Delta}{\E[\psi(\mathbb{\fsf},\Msf)]}$, which, as noted earlier, is equivalent to impose the time slot duration to be equal to the video segment duration \emph{in average}. {Note that, although not explicitly stated, the channel codeword length $n$ is upper bounded by  $\frac {D\,B\, (1- M_{u,f_u}) }{\eta_u}$.

\subsubsection{Inner source encoder} 
Generates the final multicast codeword $\bf X$ by XORing the channel codewords $\{\mathfrak{C}_u(\mathcal P_u)\}$ belonging to the same GIS, and concatenating the resulting codewords of length $n$.

Recall  that the number of GISs produced by the channel-aware source compressor is given by $\chi_{\text{CA}}$. Hence, $\bf X$ is obtained by concatenating $\chi_{\text{CA}}$ codewords of length $n$, resulting in a multicast codeword of \emph{average} length $\Delta$, as described in  Section \ref{algs}. 

%-----------------------------------------------------------------------------------------------------------------------------------------------------------------------------
% Section 6
%-----------------------------------------------------------------------------------------------------------------------------------------------------------------------------

\section{Source Compression Algorithms}
\label{algs}

In this section, we describe in detail the proposed algorithms for the channel-aware compressor. These algorithms build on generalizations of existing caching-aided coded multicast schemes to the case of noisy broadcast channels. Specifically, Section \ref{sec:CAC} describes CA-CIC as the channel aware extension of the chromatic index coding scheme [6,7,22], and Section \ref{sec:hgc} describes CA-HgC as the channel-aware extension of the polynomial-time hierarchical greedy coloring algorithm [9,11].

\subsection{Channel-Aware Chromatic Index Coding (CA-CIC)}
\label{sec:CAC}
We recall that finding a coded multicast scheme for the broadcast caching network is equivalent to finding an index code with side information given by the cache realization $\Mm$ [22]. A well-known index code is given by the chromatic number of the \emph{conflict graph}, constructed according to the demand  and cache realizations. 
Note that, given the cache realization $\Mm$, a file-level demand realization (given by the request vector ${\bf{ f}}$) can be translated into a packet-level request vector containing the missing packets at each receiver. A request for file $f_u$ by receiver $u$ is hence equivalent to requesting $D-|\Mm_{u,f}|$ packets corresponding to all missing packets of the highest quality level of video segment $f_u$. However, recall that based on the channel degradations, the sender schedules a subset of the missing packets $d_u \in \{1, \ldots , D(1-M_{u,f})\}$, as described in Section \ref{descriptions}. We denote by $\Wm$ the scheduled  packet-level configuration and by $\Wm_{u,f}$ the set of packets of file $f$ scheduled for receiver $u$. We can then define the corresponding index coding conflict graph $\mathcal{H}_{\Mm,\Wm}=(\mathcal{V}, \mathcal E)$ as follows: 
\begin{itemize}
\item Each vertex $v \in \mathcal{V}$ represents a scheduled packet, uniquely  identified by the pair  $\{ \rho(v),\mu(v)\}$,  where $\rho(v)$ indicates the  \mbox{packet identity} associated to vertex $v$ and  $\mu(v)$ represents the receiver for whom it is scheduled.  
In total, we have $|\mathcal{V}|  =\sum_{u \in \Uc } d_u$ vertices. 
\item For any pair of vertices $v_1$, $v_2$, we say that vertex $v_1$ interferes with vertex $v_2$ 
if the packet associated to the vertex $v_1$, $\rho(v_1)$, is not  in the cache of the receiver associated to vertex  $v_2$, $\mu(v_2)$, 
and $\rho(v_1)$ and $\rho(v_2)$ do not represent the same packet. Then, 
there is an edge between $v_1$ and $v_2$ in $\mathcal E$ if $v_1$ interferes with $v_2$ or $v_2$ interferes with $v_1$.
\end{itemize}

We note that in $\mathcal{H}_{\Mm,\Wm}$ the set of vertices scheduled for the same receiver, i.e., the set of vertices $v \in \Vc$ such that $\mu(v)=u$, is fully-connected. 
Based on this consideration, we refer to $\mathcal{H}_{\Mm,\Wm}$ as a $U$-clustered graph.

We now introduce the concept of Generalized Independent Set (GIS) of a $U$-clustered graph: 
\begin{defin} 
\label{def_GIS}
We define a ($s_1,\ldots, s_U$)-GIS  of a $U$-clustered graph $\mathcal{H}_{\Mm,\Wm}$ as a  set of $U$ fully connected sub-graphs $\{ \mathcal P_1, \ldots \mathcal P_U\}$ such that for all $u=1, \ldots,U$: 
\begin{itemize}
\item For all $v \in \mathcal P_u$, $\mu(v)=u$ (i.e., all the packets in $\mathcal P_u$ are scheduled for receiver $u$)
\item $|\mathcal P_u|=s_u \geq 0$ (i.e., the number of packets in $\mathcal P_u$ is equal to $s_u$)  
\item For all $i \neq u$,  $\mathcal P_u$  and $\mathcal P_i$ are mutually disconnected (i.e., no edges exist between any two subgraphs) 
\end{itemize}
\hfill $\lozenge$
\end{defin}

Note that when $s_u \leq 1, \forall u\in\mathcal U$, Def. \ref{def_GIS} becomes the classical definition of independent set. 

Based on the  notion of GIS, we introduce the definition of channel-aware valid coloring and channel-aware chromatic number of the conflict graph: 
\begin{defin} ({\bf Channel-Aware Valid Vertex-Coloring})
\label{def_coloring}
A ($\eta_1,\ldots, \eta_U$) channel-aware valid vertex-coloring is obtained by 
covering  the conflict graph $\mathcal{H}_{\Mm,\Wm}$ with ($s_1,\ldots, s_U$)-GISs  
that further satisfy $\frac{s_u}{s_i}=\frac{\eta_u}{\eta_i}, 
\,  \forall u,i\in\{1,\ldots U\}$, and assigning the same color to all the vertices in the same GIS.
\hfill $\lozenge$
\end{defin}

\begin{defin} ({\bf Channel-Aware Chromatic Number})
The ($\eta_1,\ldots, \eta_U$)  channel-aware chromatic number of a graph $\mathcal{H}$ is defined as 
\begin{align}
\label{eq:croma}
&\chi_{\text{CA}}(\mathcal H) = \min_{\{\mathcal C\}} |\mathcal C |, 
\end{align}
where $\{\mathcal C\}$ denotes the set of all  ($\eta_1,\ldots, \eta_U$) channel-aware valid vertex-colorings of $\mathcal{H}$, and $|\mathcal C |$  is the total number of colors in $\mathcal{H}$ for the given ($\eta_1,\ldots, \eta_U$) channel-aware valid vertex-coloring $\mathcal C $. 
\hfill $\lozenge$
\end{defin}

%\begin{rem}
%Computing the chromatic number of a graph is a well known NP-Hard problem.
%\end{rem}

%In the following, we provide a tight bound for the asymptotic channel-aware chromatic number $\chi_{\text{CA}}(\mathcal{H}_{\Mm,\Wm})$,  when the dimension of the graph goes to infinity, i.e. $F, D, \Delta \rightarrow  \infty$ with the ratios $B=F/D$  and  $\gamma=F/\Delta$ kept constant.

\begin{theorem}
\label{th:1}
Given a conflict graph $\mathcal{H}_{\Mm,\Wm}$ constructed according to  packet-level cache realization $\Mm$,  demand realization $\fv$, %scheduled descriptions $\{d_u=\frac {\Delta \, \eta_u}{B \, \psi(\fv,\Mm)}\}$,  
and scheduled random packet-level configuration $\Wm$, a tight upper-bound for the channel-aware chromatic number $\chi_{\text{CA}}(\mathcal{H}_{\Mm,\Wm})$,  when  $D, \Delta  \rightarrow \infty$, is given by $\psi(\mathbb{\fv},\Mm)$, i.e.,
\begin{eqnarray}
\label{eq: equality}
\chi_{\text{CA}}(\mathcal{H}_{\Mm,\Wm}) = \psi(\mathbb{\fv},\Mm) +o(1/D).
\end{eqnarray}
\end{theorem}
\hfill  $\square$
\begin{proof}
See Appendix \ref{sec: Proof of Theorem up}.
\end{proof}

As described in Section \ref{subsec:concatenated}, the GISs associated to the chromatic number $\chi_{\text{CA}}(\mathcal H)$ are then used by the concatenated source-channel encoder to generate the final multicast codeword $\Xm$.

In the following, we provide an example that illustrates the proposed multicast encoder. 

\begin{example}
\label{ex:GIS} 
{Consider a network with $U=3$ receivers, denoted by $\mathcal{U}=\{1,2, 3\}$, and $m=3$ files, denoted by $\mathcal{F}=\{W_{a}, W_b, W_c\}$. The channel rates are $\eta_1=\frac{1}{2},\, \eta_2=\frac{1}{4}, \,\eta_3=\frac{1}{4}$, and the time-slot duration is $\Delta=8$ channel uses. We assume that demand and caching distributions are such that $\E[\psi(\mathbb{\fsf},\Msf)]=2$. 
Each file is multiple description coded into $D$ descriptions, e.g., $W_a=\{ W_{a,1}, W_{a,2}, \ldots, W_{a,D}\}$, each of length $B=F/D=1$ bit.}
According to Eq. \eqref{eq:descriptionsAv}, the sender schedules 2 descriptions for  receiver 1, i.e., $d_1=2$,  and one description for receivers 2 and 3, i.e., $d_2=d_3=1$. 
{We assume the caching realization is given by: $\Mm_{1, W_a}=\{ W_{a,1}, W_{a, 2}\}$, $\Mm_{1,W_c}=\{W_{c,1}\}$; $\Mm_{2,W_b}=\{W_{b,1}, W_{b,2}\}$, $\Mm_{2, W_c}=\{W_{c,1}\}$; $\Mm_{3, W_a}=\{ W_{a,1}\}$, $\Mm_{3,W_b}=\{W_{b,1}, W_{b,3}\}$. 
Suppose receiver 1 requests $W_b$, receiver 2 requests $W_a $, and receiver 3 requests $ W_c$ such that $\Wm_{1,W_b}=\{W_{b,1}, W_{b,2}\}$, $\Wm_{2,W_a}= \{W_{a,1}\}$ and $\Wm_{3, W_c}= \{W_{c,1}\}$. }
The corresponding conflict graph $\mathcal{H}_{\Mm, \Wm}$ is shown in Fig.~\ref{fig_ex1}, along with the U-clusters $\{\Vc_u\}_{u=1}^3$. 
Each vertex $v$ in $\mathcal{H}_{\Mm, \Wm}$ is uniquely  identified by the pair  $\{ \rho(v),\mu(v)\}$  with $\rho(v)$ indicating the \mbox{packet identity} and $\mu(v)$ the \mbox{requesting receiver}.

\begin{figure}[t]
\centering
\includegraphics[width=0.7\linewidth]{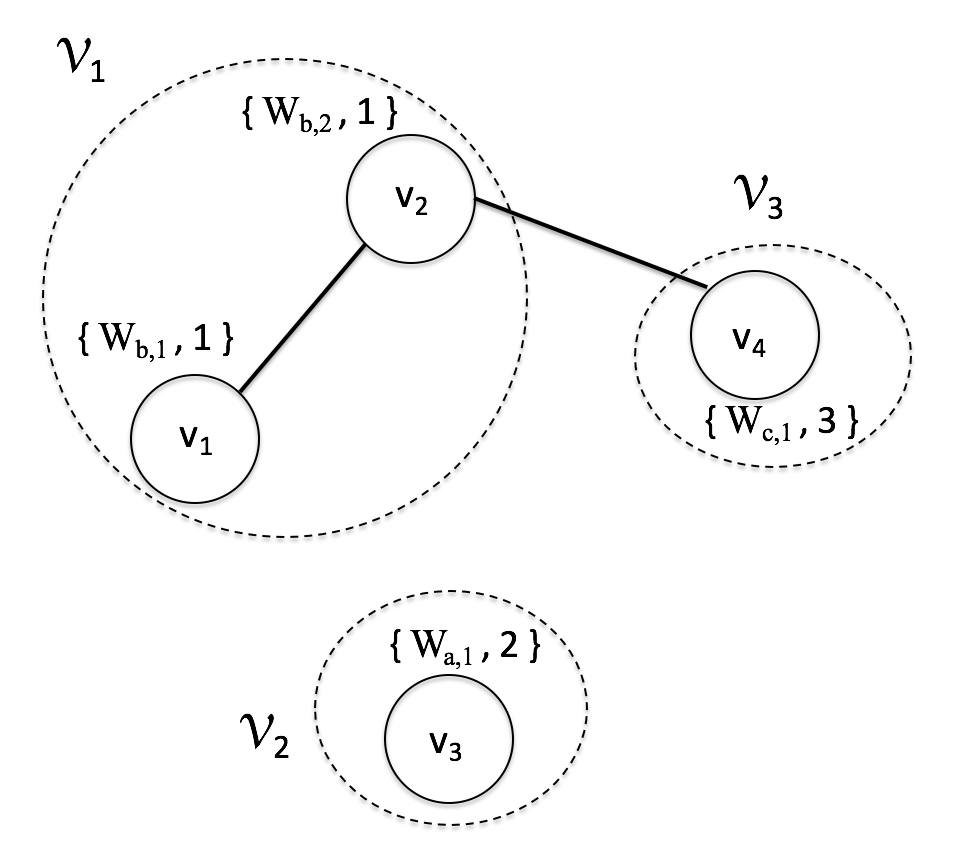}
\caption{An example of $U$-clustered graph $\mathcal{H}_{\Mm,\Wm}$ }
	\label{fig_ex1}
\end{figure}

\begin{figure}[t]
\centering
\includegraphics[width=0.75\linewidth]{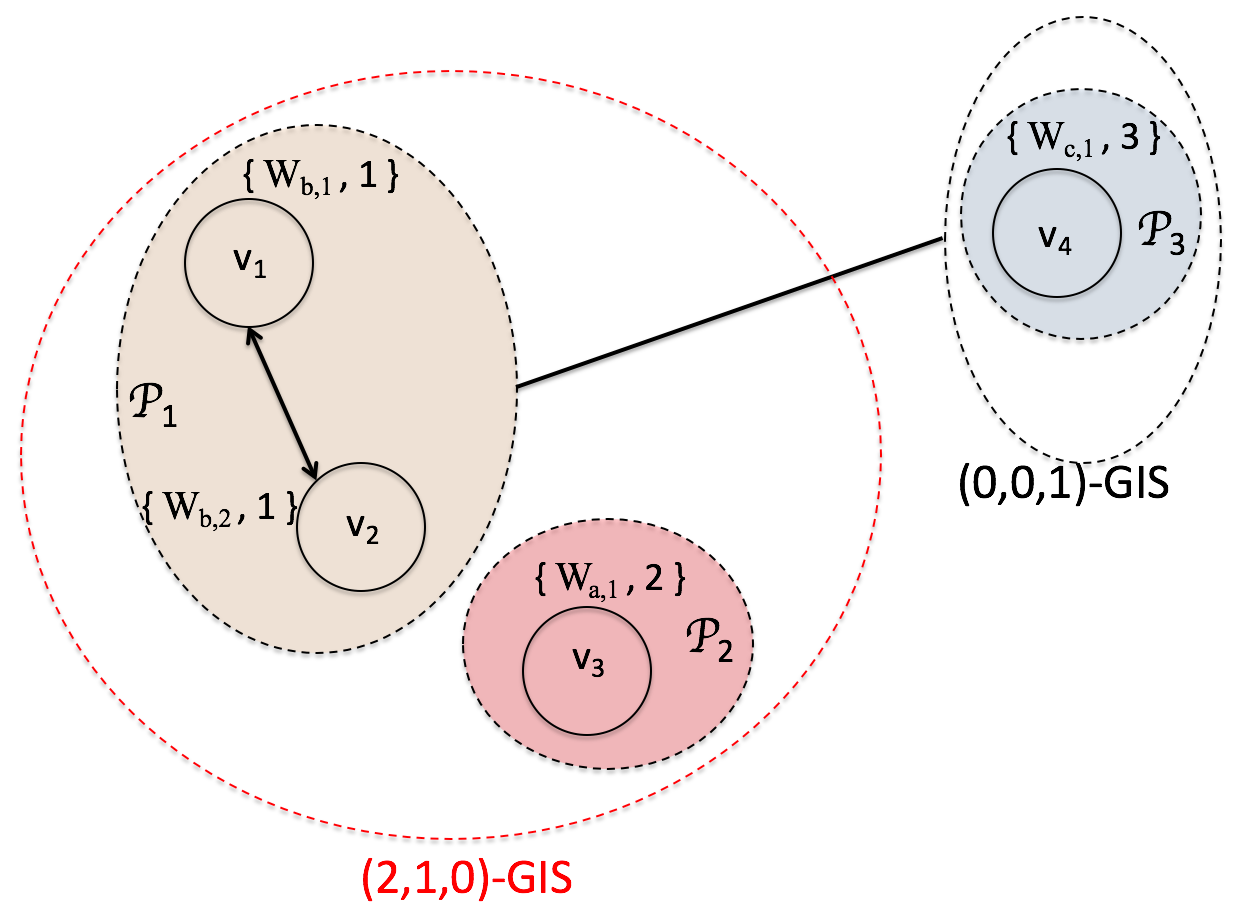}
\caption{An example of ($s_1,s_2, s_3$)-GISs  of the $3$-clustered graph $\mathcal{H}_{\Mm,\Wm}$ shown in Fig.~\ref{fig_ex1}.}
	\label{fig_gis}
\end{figure}

In Figure~\ref{fig_gis}, we show two ($s_1,s_2, s_3$)-GISs covering the $3$-clustered graph $\mathcal{H}_{\Mm, \Wm}$. The first GIS is composed of  two nonempty fully connected sub-graphs: $\mathcal{P}_1=\{v_1, v_2\}$ with $s_1=|\mathcal P_1|=2$, and $\mathcal{P}_2=\{v_3\}$ with $s_2=|\mathcal P_2|=1$. The second GIS is composed of one non-empty fully connected subgraph: $\mathcal{P}_3=\{v_4\}$, with $s_3=|\mathcal P_3|=1$.

The channel-aware vertex coloring is depicted in Fig. \ref{fig_gis}, with channel-aware chromatic number $\chi_{\text{CA}}(\mathcal{H}_{\Mm,\Wm})$ equal to two colors, one for each GIS (red dotted circle and black dotted circle). The multicast codeword $\Xm$ of length $\Delta=8$ is obtained concatenating the codewords $\mathfrak C_1(\mathcal P_1) \oplus\mathfrak C_2(\mathcal P_2)$ and $\mathfrak C_3(\mathcal P_3)$, each of length $n=4$.
\end{example}

It is important to note that computing the chromatic number of a graph is a well known NP-Hard problem. In the next subsection, we introduce a polynomial-time approximation of CA-CIC, referred to as {\em Channel-Aware Hierarchical greedy Coloring (CA-HgC)}.

\subsection{Channel-Aware  Hierarchical greedy Coloring (CA-HgC)}
\label{sec:hgc}

We now present %In this section, motivated by the exponential complexity of CA-CIC, we present 
Channel-Aware Hierarchical greedy Coloring (CA-HgC), a novel coded multicast algorithm that fully accounts for the broadcast channel impairments, while exhibiting polynomial-time complexity. 

The CA-HgC algorithm works by computing two valid colorings of the conflict graph $\mathcal{H}_{\Mm,\Wm}$, referred to as CA-HgC$_1$ and CA-HgC$_2$. CA-HgC then compares the corresponding number of colors achieved by the two solutions and selects the coloring with minimum number of colors. We note that CA-HgC$_1$ coincides with the conventional naive (uncoded) multicast scheme. In fact, CA-HgC$_1$ computes the minimum coloring of $\mathcal{H}_{\Mm,\Wm}$ subject to the constraint that only the vertices representing the same packet are allowed to have the same color. In this case, the total number of colors is equal to the number of distinct requested packets, and the coloring can be found in $O(|\mathcal{V}|^2)$. On the other hand, CA-HgC$_2$ is described by Algorithm~\ref{alg:1}, in which the subroutine { \tt  GISfunction}$(\cdot,\cdot,\cdot)$ is defined by Algorithm~\ref{alg:2}. It can be shown that the complexity of CA-HgC$_2$ is given by $O(U |\mathcal{V}|^2)$, i.e., it is polynomial in $|\mathcal{V}|$.

\begin{algorithm}[b!]
\label{alg_hg}
\caption{CA-HgC$_2$}
\label{alg:1}
\begin{algorithmic}[1]
\STATE $\mathcal{C} = \emptyset$
\STATE ${\bf{X}} = \emptyset$
\FOR{$i=U:1$}
\FOR{$v \in H_i : |\Kc_v| = \min_{v' \in H_i} \{ | \Kc_{v'} | \}$}
\STATE $[\mathcal{G}, \Omega] = ${\tt  GISfunction}$(H_i,v,i)$\\
\IF{$\mathcal{G} \neq \emptyset$}
\STATE $\forall u \in \Omega$ code the vertices in $\{ v' \in \mathcal{G} : \mu(v') = u\}$ with the $\left(\frac{\Delta}{\E[\psi(\mathbb{\fsf},\Msf)]}\right)$--length codeword $ \mathfrak c_u=\mathfrak{C}_u(\{ v' \in \mathcal{G} : \mu(v') = u\})$ in the codebook $\mathfrak{C}_u$
\STATE ${\bf{X}} = [{\bf{X}}\,\, \sum_{u \in \Omega}\,\,\otimes \,\, \mathfrak c_u]$
\STATE color $(\sum_{u \in \Omega}\,\,\otimes \,\, \mathfrak c_u)$ by $\alpha \not\in \mathcal{C}$
\STATE $ H_i = H_i \setminus \mathcal{G}$
\ELSE
\STATE $ H_i = H_i \setminus \{v\}$; $H_{i-1}=H_{i-1} \cup \{ v\}$
\ENDIF
\ENDFOR
\ENDFOR
\RETURN $|\mathcal{C}|$, ${\bf{X}}$
\end{algorithmic}
\end{algorithm}

\begin{algorithm}[b!]
\caption{{ \tt  GISfunction}$(H_i,v,i)$}
\label{alg:2}
\begin{algorithmic}[1]
\STATE $\mathcal{G} = \{ v \}$; $\Omega = \{ \mu(v) \}$; $\tilde{H_i} = H_i$
\FOR{$v' \in \tilde{H_i} \setminus \mathcal{G} : |\Kc_{v'}| = \min_{v'' \in \tilde{H_i} \setminus \mathcal{G}} \{ |\Kc_{v''}| \}$}
\IF{no edge between $v'$ and $\mathcal{G}$}
\STATE $\mathcal{G} = \mathcal{G} \cup \{v'\}$; $\Omega = \Omega \cup \{\mu(v')\}$
\ELSE
\STATE $\tilde{H_i} = \tilde{H_i} \setminus \{v'\}$
\ENDIF
\ENDFOR
\IF {$|\mathcal{G}| \geq i$}
\FOR{$v' \in \tilde{H_i} \setminus \mathcal{G} : |\Kc_{v'}| = \min_{v'' \in \tilde{H_i} \setminus \mathcal{G}} \{ |\Kc_{v''}| \} \wedge \mu(v') \in \Omega$}
\STATE $\mathcal{G}_{\mu(v')} =  \{ v" \in \mathcal{G} : \mu(v") = \mu(v') \}$
\IF{$|\mathcal{G}_{\mu(v')}| < d_{\mu(v')}$ and no edge between $v'$ and $\mathcal{G} \setminus \mathcal{G}_{\mu(v')}$}
\STATE $\mathcal{G} = \mathcal{G} \cup \{v'\}$
\ELSE
\STATE $\tilde{H_i} = \tilde{H_i} \setminus \{v'\}$
\ENDIF
\ENDFOR
\RETURN $\mathcal{G}$, $\Omega$
\ELSE
\RETURN $\mathcal{G} = \emptyset$, $\Omega = \emptyset$
\ENDIF
\end{algorithmic}
\end{algorithm}

In the following, we first guide the reader through Algorithm~\ref{alg:1} and then we provide an illustrative example\footnote{With a slight abuse of notation, we denote by $\mathfrak c_u=\mathfrak{C}_u(\{ v' \in \mathcal{G} : \mu(v') = u\})$ the codeword obtained by coding the $B s_u$ source bits associated to the packets $\{\rho(v'): \mu(v') = u\}$. }. 

Let $\mathcal{K}_v \eqdef \{ u \in \mathcal{U}: \rho(v) \in \Wm_u \cup \Mm_u\}$ denote the set of receivers that request and/or cache packet $\rho(v)$. We start by initializing the $i$-th vertex hierarchy in of $\mathcal{H}_{\Mm,\Wm}$ to the set of vertices for whom the number of receivers requesting and/or caching its packet identity is equal to $i$, i.e., $\mathcal{H}_i \eqdef \{v : |\mathcal{K}_v|=i\}$. CA-HgC$_2$ proceeds in decreasing order of hierarchy starting from the $U$-th hierarchy.

For each vertex $v$ in the $i$-th hierarchy in increasing order of cardinality $|\mathcal{K}_v|$, we call the subroutine { \tt GISfunction}$(\mathcal H_i,v,i)$. If the subroutine returns a non-empty set $\mathcal{G}$, then we color the vertices in $\mathcal{G}$ with the same color (lines 6-10). Otherwise, we move the uncolored vertex $v$ to the next hierarchy, i.e., $\mathcal{H}_{i-1}$ (lines 11-12). This procedure is iteratively applied for any hierarchy, until all the vertices in the conflict graph $\mathcal{H}_{\Mm, \Wm}$ are colored. 
The procedure returns the number of {colors $|\mathcal{C}|$} as well as the set of codewords ${\bf{X}}$. 

Regarding the subroutine { \tt  GISfunction}$(\mathcal H_i,v,i)$ shown in Algorithm~\ref{alg:2}, we initially set $\mathcal G = \{v\}$.  Then, for each vertex $v'\in\mathcal H_i$ in increasing order of cardinality $|\mathcal K_{v'}|$, we move $v'$ into $\mathcal G$ if $v'$ is not connected (via the conflict graph) to any node currently in $\mathcal G$ (lines 2-8). Note that in the first iteration the lowest cardinality in $\mathcal{H}_i$ is exactly equal to $i$. By construction, each vertex in the conflict graph has an associated receiver, hence we denote by $\Omega$ the set of receivers associated to the vertices in $\mathcal G$. If $\mathcal{G} < i$, i.e., if we were not able to select at least $i$ vertices, then we return two empty sets (lines 19-20). Otherwise, we proceed by trying to select, for each receiver $\mu(v') \in \Omega$, at most $d_{\mu(v')} -1$ additional lowest-cardinality vertices in $\mathcal{H}_i$ having no links with the vertices in $\mathcal G$ associated to receivers different from $\mu(v')$ (lines 10-17). Once this is done (or if the considered hierarchy is 1, meaning that we have no further hierarchies to explore), then we return the set of selected vertices $\mathcal{G}$ as well as the set of associated receivers $\Omega$ (line 18).

\begin{example} Consider the same conflict graph $\mathcal{H}_{\Mm, \Wm}$ shown in Fig.~\ref{fig_ex1} and let us apply Algorithm~\ref{alg:1}. We start by constructing $\mathcal{K}_v$ by inspection of the conflict graph: $\mathcal{K}_{v_1}=\{1, 2, 3\}$, $\mathcal{K}_{v_2}=\{1, 2\}$, $\mathcal{K}_{v_3}=\{1, 2, 3\}$, and $\mathcal{K}_{v_4}=\{1, 2, 3\}$. Hence, $\mathcal{H}_U=H_3=\{v_1, v_3, v_4\}$, $\mathcal{H}_2 = \{ v_2\}$, and $\mathcal{H}_1 = \emptyset$. Starting from the highest hierarchy, in line 5 of Algorithm~\ref{alg:1} we call the subroutine \textit{GISfunction} with parameters $\mathcal{H}_3$, $v_1$ and $3$. By running iteratively the proposed algorithm, the GISs shown in Fig.~\ref{fig_gis} are obtained.
\end{example}

%-----------------------------------------------------------------------------------------------------------------------------------------------------------------------------
% Section 7
%-----------------------------------------------------------------------------------------------------------------------------------------------------------------------------

\section{Multicast Decoder }
\label{sec: specialdec}

In this section, we analyze the decodability of the proposed achievable scheme.

From the observation of its channel output ${\bf Y}_u$, representing its noisy version of the transmitted codeword ${\bf X}$,
receiver $u$ {decodes} the $d_u$ descriptions  of its requested video {segment} $W_u \in \mathcal F$  scheduled and transmitted by the sender as  $\widehat{W}_u = \lambda_u(\mathbf{Y}_u,Z_u,\mathbf{f})$, via its own decoding function $\lambda_u$, which consists of two stages:
\begin{itemize}
\item First, receiver $u$ is informed (e.g.,  via packet header information) of the sub-codewords in the concatenated multicast codeword $\bf X$ that contain any of its scheduled descriptions. 
For each of the identified sub-codewords, receiver $u$ obtains 
the noisy version of its channel codeword {$\mathfrak{C}_u(j)$ with $j \in \{1, \ldots,2^{n  \eta_u} \}$} by performing the inverse XOR function. %This processing can be conducted since, by design, the codewords of different receivers belonging to the same GIS do not interfere with each other (i.e., they can be generated from the cached information of receiver $u$). In fact, 
To this end, receiver $u$ is informed of %$i)$ all the $U$ codebooks; $ii)$ 
{the packets and their intended receivers that are present in the XORed sub-codeword.
%along with the corresponding receiver codebooks.
 } % used to construct the corresponding receiver codewords.}
Receiver $u$ can then construct the channel codewords associated to the other receivers from its cached information and the corresponding codebooks (recall that every receiver is informed of all the channel codebooks) and recover its own codeword $\mathfrak{C_u(j)}$ via inverse XORing. % with the received multicast codeword.
\item Then, the recovered noisy codeword {$\mathfrak{C}_u(j)$} is sent to the channel decoder of receiver $u$, which reconstructs the bits associated to the set $\mathcal P_u$ according to the channel rate $ \eta_u$.
\end{itemize}
Hence, according to Def.~\ref{def:admissible}, in the limit $D \rightarrow \infty$, the proposed sequence of $\left( \Delta,  \{\epsilon_u\} \right)$-delivery schemes is {\em admissible.}

\subsection{Achievable Delivery Rates}
\label{sec:4AchivDist}
It is immediate to verify that the $\left( \Delta,  \{\epsilon_u \} \right)$-delivery schemes RAP-CA-CIC  and RAP-CA-HgC {generate} a sequence of admissible schemes as per Def.  \ref{def:admissible}. In the following, we provide an explicit expression for the average per-receiver and network delivery rates, formally defined in Eqs. \eqref{average-rate-user}-\eqref{average-rate}, achieved by the  proposed $\left( \Delta,  \{\epsilon_u \} \right)$-delivery schemes {when $F, D, \Delta \rightarrow  \infty$ with the ratios $B=F/D$  and $\gamma=\Delta/F$ kept constant. For ease of exposition, we assume $p_{f,u} = p_f$, $q_{f,u} = q_f$, and $M_u=M, \, \forall u \in \mathcal U$.\footnote{The generalization to the case of different $\{p_{f,u}\}$, $\{q_{f,u}\}$ and $\{M_u\}$  across receivers is immediately obtained by replacing Eqs.~\eqref{eq:psi}-\eqref{eq:rho} with Eqs.~(2)-(3) in \cite{ji2015Preserving}.}

\begin{theorem}
\label{theorem: 1}
The average per-receiver and network delivery rates achieved by the $\left( \Delta,  \{\epsilon_u\} \right)$-delivery schemes RAP-CA-CIC  and RAP-HgCAC,
as $F, D, \Delta \rightarrow \infty$,  satisfy: % \textcolor{red}{the inequalities \eqref{eq:ratefinal} and \eqref{eq:ratefinal_bis}, respectively}:
\begin{align}
& R_u^{\rm CA-CIC}(U,m,M,\{\epsilon_u\}) \geq  R_u^{\rm CA-HgC}(U,m,M,\{\epsilon_u\}) \nonumber \\
& \qquad\qquad \geq \min \left \{  \frac{ \eta_u}{\E[\psi(\mathbb{\fsf},\Msf)] } , \frac{(1- \E[M_{u,\fsf_u}])}{\gamma}   \right \} + \frac{\E[M_{u,\fsf_u}]}{\gamma},
\label{eq:ratefinal}
\end{align}
\begin{align}
\label{eq:ratefinal_bis}
& R^{\rm CA-CIC}(U,m,M,\{\epsilon_u\})  \geq    R ^{\rm CA-HgC}(U,m,M,\{\epsilon_u\}) \nonumber\\
& \quad \geq \sum_{u=1}^U \! \left (\min\left\{ \frac{ \eta_u}{\E[\psi(\mathbb{\fsf},\Msf)] } , \frac{(1- \E[M_{u,\fsf_u}])}{\gamma}  \right\}  +  \frac{\E[M_{u,\fsf_u}]}{\gamma} \right), 
\end{align}
where
\begin{align}
\label{eq:chi}
&\qquad \E[\psi(\mathbb{\fsf},\Msf)] ={\min\{\phi(\pv,\Qm),\bar m\}}, 
\end{align}
with
\begin{align}
& \bar m = \sum_{f=1}^m \left(1 - \left(1 - q_f\right)^{U} \right), \nonumber \\
& \phi(\pv,\Qm) = \sum_{f=1}^m \sum_{\ell=1}^U {U \choose \ell}  \rho_{f,\ell} (1-p_{f} M)^{U-\ell+1} (p_{f} M)^{\ell-1}, \label{eq:psi} \\
& \displaystyle \rho_{f,\ell} \eqdef  \mathbb \PP(f = \arg\! \max_{j \in \Upsilon} \,\,\, (p_{j}M)^{\ell-1}(1-p_{j}M)^{n-\ell+1}) ,\label{eq:rho}
\end{align}
and $\Upsilon$ denoting a random set of $\ell$ elements selected in an i.i.d. manner from $\mathcal F$ (with replacement).
\hfill  $\square$
\end{theorem}
\begin{proof}
See Appendix \ref{sec: Proof of Theorem up}.
\end{proof}

%\textcolor{red}{Note that in Theorem \ref{theorem: 1} we have used that $\Delta=F\gamma$
% Denoting by $R_{pb}$ the payback rate of each  segment and denoting by $Q_b$ the buffer size in video segment at each receiver it follows that $\Delta R_{pb} =Q_b *F$. Hence  $$\frac{ B \E[\mu_{u,\fsf_u}] }{\E[t_{u}(fsf)] }  \geq \frac{B \E[\mu_{u,\fsf_u}] }{\Delta}= \frac{F \E[M_{u,\fsf_u}] }{\Delta} =\frac{R_{pb} \E[M_{u,\fsf_u}] }{Q_b}$$    
%}

\begin{rem}  
Observe from Eq. \eqref{eq:ratefinal} that, in contrast with the classical SSC-CC solution, our proposed channel-aware caching-aided coded multicast solutions guarantee that no receiver gets affected by receivers with worse channel conditions.
\end{rem}

%-----------------------------------------------------------------------------------------------------------------------------------------------------------------------------
% SECTION 8
%-----------------------------------------------------------------------------------------------------------------------------------------------------------------------------
\section{Validation of the Theoretical Results}
\label{sec:6}
The goal of this  section is provide numerical validation of the theoretical results  given in Section~\ref{sec:4AchivDist}. 
Specifically, we analyze the performance of RAP-CA-HgC for finite file packetization (i.e., finite number of video descriptions) compared with: \emph{i)} LFU caching with Compound Channel transmission (i.e., naive multicasting at the rate of the worse channel receiver), referred to  as LFU-CC; \emph{ii)} LFU caching with unicast transmission, referred to as orthogonal LFU (O-LFU); \emph{iii)} RAP caching with separate source-channel coding over compound channel with HgC as source encoder, referred to as RAP-SSC-CC; and \emph{iv)} the benchmark upper bound (RAP-CA-HgC) when $D=\infty$ (Theorem~\ref{theorem: 1}). 

We consider a network with $m =1000$ files and $U=30$ receivers with channel rates uniformly distributed among three values usually used in LTE standards: $\{\eta_1=\frac{1}{2}, \eta_2=\frac{3}{4}, \eta_3=\frac{1}{4}\}$. We assume that receivers request files according to a common Zipf demand distribution with $\alpha=\{0.2, 0.4\}$, all caches have size $M$ files, and the maximum number of descriptions is $D=200$. 

Figs. \ref{fig_col1} and \ref{fig_col2} show numerical results in terms of \emph{network load}, defined as the inverse of the {\em network transmission rate},  and given by $\frac{\E[\psi(\mathbb{\fsf},\Msf)]}{\sum_{u=1}^U \E[\eta_u( \mathbb{\fsf})]}$, as shown in Theorem \ref{theorem: 1}. This metric is specially illustrative of the the amount of bandwidth resources the wireless operator needs to provide in order to meet the receiver demands. 

In Fig.~\ref{fig_col1}, we assume a Zipf parameter $\alpha=0.2$. Observe first the performance of LFU. As expected, the load reduces approximately linearly with the cache size $M$. Note that while a channel-unaware caching-aided coded multicast scheme such as RAP-SSC-CC is able to achieve smaller network loads than LFU for all values of $M$, the reduction is limited by the adverse effect of the heterogeneous channel conditions on the coded multicast transmission. In fact the performance of RAP-SSC-CC is essentially the same as that of LFU-CC, clearly showing the vanishing benefit of coded multicasting under heterogeneous channel conditions. 
Observe now the performance of the proposed RAP-CA-HgC with $D=\infty$. The reduction of network load with cache size resembles the remarkable multiplicative caching gains of the caching-aided coded multicast schemes under error-free broadcast channel [6,7,8]. This clearly shows the effectiveness of RAP-CA-HgC in allowing receivers to decode their requested video segments without being affected by receivers with worse channel conditions. 
Note that RAP-CA-HgC achieves network load reductions that are $7\times$ larger than LFU for $M=200$ (20\% of the library size).
Importantly, the more practical RAP-CA-HgC with $D=200$ is able to preserve the multiplicative rate-memory tradeoff, with a notorious $4\times$ load reduction with respect to LFU for $M=200$. 

 Note that very similar trends are observed in Fig. \ref{fig_col2}, where the Zipf parameter is set to $\alpha=0.4$, confirming the superiority of RAP-CA-HgC in exploiting the remarkable benefits of caching-aided coded multicast video delivery schemes, while avoiding critical penalizations from the presence of receivers with worse channel conditions. 

\begin{figure}[t]
\centering
\includegraphics[width=0.85\linewidth]{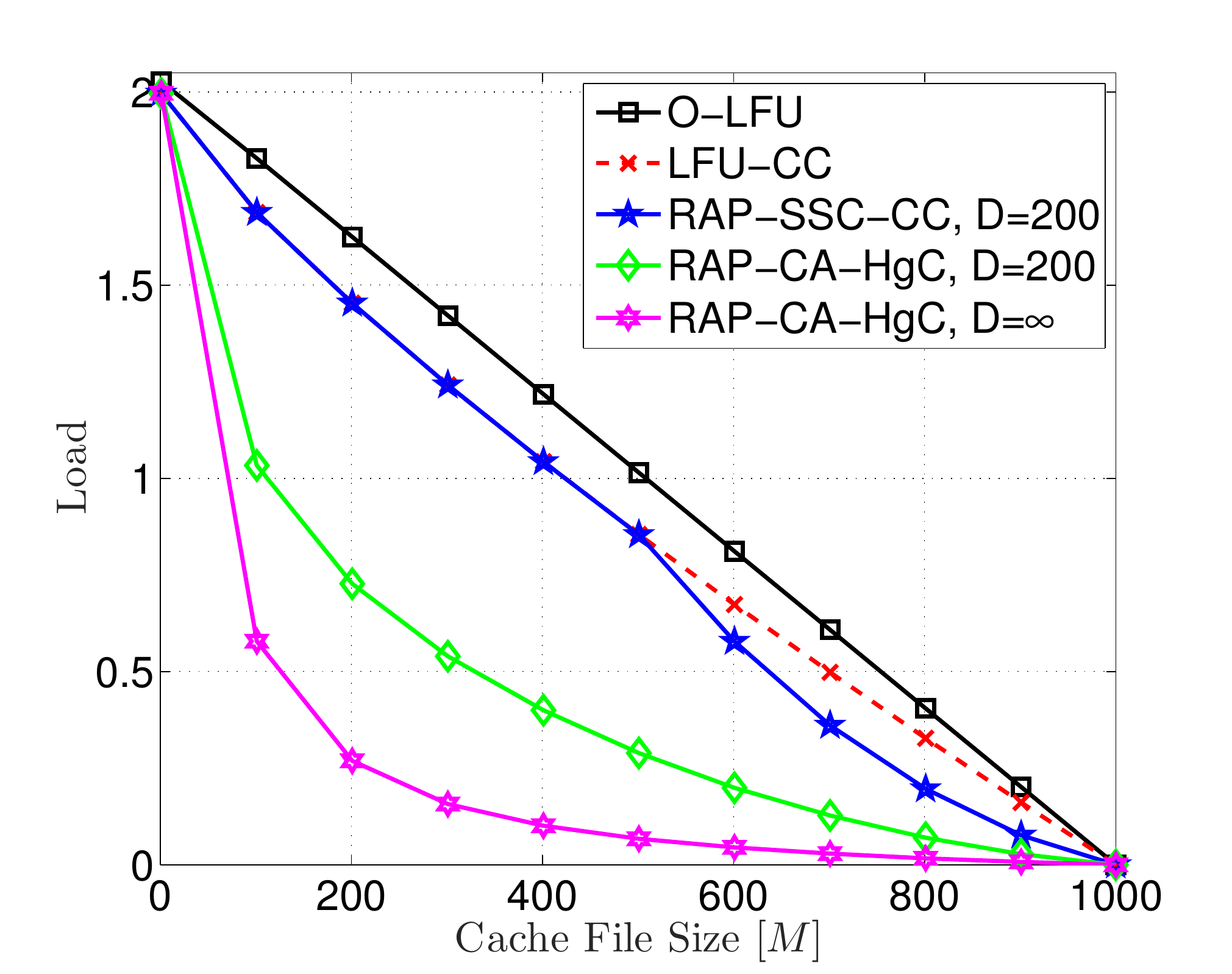}
\caption{Average load in a broadcast caching network with $m=1000, U=30, \alpha=0.2$.}
	\label{fig_col1}
\end{figure}

\begin{figure}[t]
\centering
\includegraphics[width=0.85\linewidth]{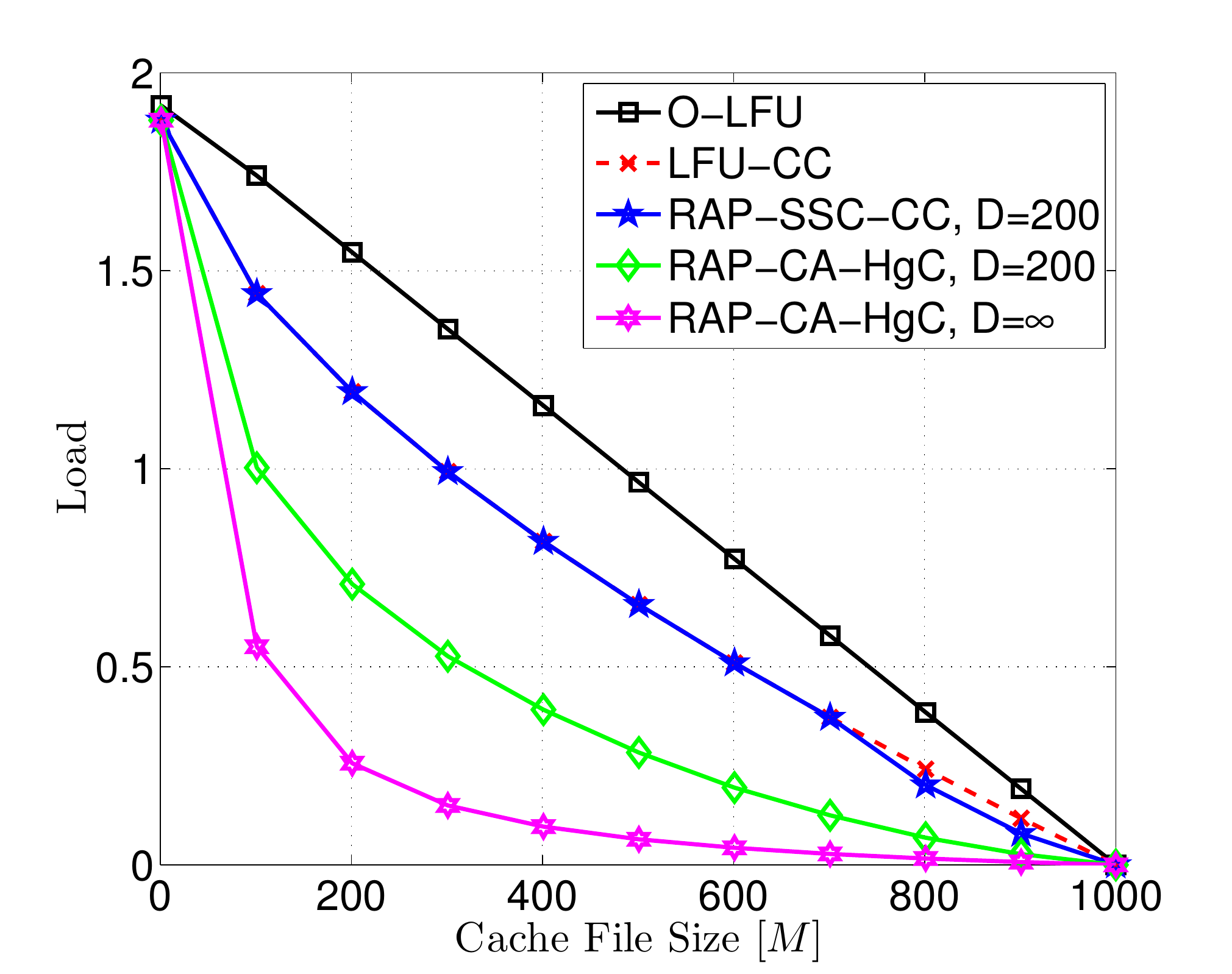}
\caption{Average load in a broadcast caching network  with $m=1000, U=30, \alpha=0.4$.}
	\label{fig_col2}
\end{figure}

%-----------------------------------------------------------------------------------------------------------------------------------------------------------------------------
% SECTION 9
%-----------------------------------------------------------------------------------------------------------------------------------------------------------------------------

\section{Conclusions}
\label{sec:7}
In this paper, we design and analyze a novel wireless video delivery scheme that specifically addresses the pressing need to accommodate next-generation video-based services over increasingly crowded and bandwidth-limited wireless access networks. 
We %\DIFdelbegin \DIFdel {extends the recently proposed RAP-CIC scheme [7] by considering}\DIFdelend \textcolor{red}{
consider a heterogeneous wireless edge caching architecture where the cache-enabled receivers (e.g., helpers) are connected to the sender by lossy links.
We provide an information theoretic formulation for the channel aware caching-aided coded multicast problem and design two achievable schemes, RAP-CA-CIC and RAP-CA-HgC, which result from the careful implementation of joint source-channel coding into the caching-aided coded multicast problem. 
Our solutions {preserve the cache-enabled multiplicative throughput gains of the uniform channel scenario, 
%guarantee the highest admissible video delivery rate to each receiver for the given propagation conditions, i.e., 
completely avoiding throughput penalizations from the presence of receivers experiencing worse propagation conditions.}

%-----------------------------------------------------------------------------------------------------------------------------------------------------------------------------
% SECTION 10
%-----------------------------------------------------------------------------------------------------------------------------------------------------------------------------
\appendices
\section{Proof of Theorems \ref{th:1} and \ref{theorem: 1}}
\label{sec: Proof of Theorem up} 

In the following, we analytically quantify the performance of the RAP-CA-HgC and, by extension, of RAP-CA-CIC. 

For mathematical tractability, we %derive an \red{upper bound} by 
relax RAP-CA-HgC by removing the conditions in line 9 of Algorithm~\ref{alg:2}. 
This forces the construction of a GIS to happen within a given hierarchy, without taking into account the uncolored vertices from higher hierarchies.\footnote{This in general returns a less efficient covering of the graph, i.e., a larger number of disjoint GISs covering the graph. However, as proved in \cite{ji2015random} when $D \rightarrow \infty$ this provides a tight upper bound.}
This implies that any GIS, $\mathcal G$, generated by Algorithm~\ref{alg:1} is composed of vertices belonging to the same hierarchy, i.e., $\forall v, v' \in \mathcal G, \Kc_v=\Kc_{v'}$.
This allows us to associate any GIS, $\mathcal G$, generated by  Algorithm~\ref{alg:1} and belonging to the $\ell$-th  hierarchy with a given subset of $\ell$
receivers, $\Uc^\ell$, such that $\Kc_v=\Uc^\ell$ for any $v \in \mathcal G$.

Let  $\Jc(\Uc^\ell)$ be the ($s_1,\ldots, s_U$)-GIS associated with $\Uc^\ell$, i.e., the GIS for which $s_u =0$  if $u \notin \Uc^\ell$, with $\ell=1, \ldots, U$. Denoting by $|\Jc(\Uc^\ell)|$ the number of coded bits associated with $ \Jc(\Uc^\ell)$, we have
\begin{equation}
\label{eq:cardi}
 |\Jc(\Uc^\ell)|=   \max_{ f \in \fv(\mathcal U^\ell)}  \left \{ \left \lceil   \frac{B}{ \eta_{\mu(v)}} \sum_{v:  \rho(v)  \ni  f } 1\{\Kc_{v} =\Uc^\ell  \} \right \rceil\right \}, 
\end{equation}
where $\fv(\mathcal U^\ell)$ represents the set of files requested by $\Uc^\ell$ and the indicator $1\{\Kc_{v} = \Uc^\ell  \}$ denotes the event that packet $\rho(v)$ requested by receiver ${\mu(v)}  \in \Uc^{\ell}$ is cached at all receivers $\Uc^\ell \setminus \{\mu(v)\}$ and not cached at any other receiver. 
Note that  $1\{\Kc_{v} =\Uc^\ell \}$ follows a Bernoulli distribution with parameter \begin{eqnarray}
\lambda(f,\ell )&=& (p_{f}M)^{\ell -1}  (1-p_{f}M)^{U -\ell +1}.
\end{eqnarray}
Denoting by $d_{\mu(v)} $ the scheduled descriptions for receiver $\mu(v)$, and by exploiting the Bernoulli distribution of $1\{\Kc_{v} =\Uc^\ell \}$, with high probability, we have
\begin{equation}
\label{eq:bern}
\sum_{v:  \rho(v)  \ni  f_{\mu(v)} }1\{\Kc_{v} = \Uc^\ell\} = \lambda(f,\ell )d_{\mu(v)}   + \delta(D), 
\end{equation}
with $ \delta(D) \rightarrow 0$ as $D \rightarrow \infty $.  

Substituting \eqref{eq:bern} in \eqref{eq:cardi}, we obtain
\begin{eqnarray}
\label{eq: cliquedoded}
|\Jc(\Uc^\ell)| 
\!\!\! \!\!\!  & = & \!\!\!  \!\!\! \max_{ f \in \fv(\mathcal U^\ell)}  \left \{ \left \lceil  \frac{B}{  \eta_{\mu(v)}} \left(  \lambda(f,\ell )d_{\mu(v)}   + \delta(D) \right) \right \rceil\right\}. 
\notag\\
\end{eqnarray}

By assumption (see Eq. \eqref{eq:descriptions}), we set $d_{\mu(v)} $, which represents the cardinality of the set $\{v:  \rho(v)  \ni  f_{\mu(v)}\}$, such that 
\begin{eqnarray}
\label{eq:d_uproof}
\frac { B}{\eta_{\mu(v)}} d_{\mu(v)}  = \min \left \{ \frac {\Delta}{\psi(\fv,\Mm) } ,  \frac { B D(1-p_{f_u} M ) }{\eta_{\mu(v)}}   \right  \} .
\end{eqnarray}

If $\frac {\Delta}{\psi(\fv,\Mm) }  < \frac { B D(1-p_{f_u} M ) }{\eta_{\mu(v)}}$, then, 
from \eqref{eq:d_uproof}, it also follows that if the channel codeword  length is given by $n= \frac {\Delta}{\psi(\fv,\Mm) }$, then only  one channel codeword per receiver is associated to each GIS. 
Hence, from the scheme described in Section \ref{subsec:concatenated},  the number of codewords that the inner source encoder concatenates is equal to the number of GISs, and the total number of coded bits is 
\begin{eqnarray}
\label{eq:czz}
\Delta= n \, \chi_{\text{CA}}(\mathcal{H}_{\Mm,\Wm}). 
\end{eqnarray}

By substituting \eqref{eq:d_uproof} in \eqref{eq: cliquedoded}, we have
\begin{eqnarray}
\label{eq: cliquedoded1}
|\Jc(\Uc^\ell)| 
\!\!\! \!\!\!  &=& \!\!\! \!\!\!
\max_{ f \in \fv(\mathcal U^\ell)}  \left \{ \left \lceil  \frac{\Delta }{ \psi(\fv,\Mm) } \left(  \lambda(f,\ell )   + o(1/D) \right) \right \rceil\right\}.
\notag
\end{eqnarray}
As $D \rightarrow \infty$,  with high probability,
\begin{eqnarray}
\label{eq: clique 1}
&& |\Jc(\Uc^\ell)| =  \max_{f_u \in \fv(\mathcal U^\ell)} \quad \frac{\Delta}{ \psi(\fv,\Mm)}\lambda(f,\ell), 
\end{eqnarray}
from which the total number of coded bits is given by
\begin{eqnarray}
\label{eq: chromatic number 1}
\sum_{\ell=1}^U \sum_{\Uc^\ell \subset \Uc}\left|\Jc(\Uc^\ell)\right|
=  \frac{\Delta}{\psi(\fv,\Mm)} \sum_{\ell=1}^U  { U \choose  \ell} \max_{f_u \in \fv(\mathcal U^\ell)}\lambda(f,\ell).
\end{eqnarray}
From \eqref{eq: chromatic number 1}, in order to satisfy the time-slot duration constraint $\Delta$, we need
\begin{eqnarray}
\label{eq: chromatic number 2}
\psi(\fv,\Mm)= \sum_{\ell=1}^U   { U \choose  \ell} \max_{f_u \in \fv(\mathcal U^\ell)}   \lambda(f,\ell).
\end{eqnarray}
Furthemore, from \eqref{eq:czz} and \eqref{eq: chromatic number 1},  it follows that $\psi(\fv,\Mm) = \chi_{\text{CA}}(\mathcal{H}_{\Mm,\Wm}) $ from which Theorem \ref{th:1} follows. By averaging  \eqref{eq: chromatic number 2} over the demand distribution, we obtain that  $\psi(\fsf,\Msf) $ concentrates its probability mass  in   $\E[ \psi(\fv,\Mm)] $ with
\begin{eqnarray}
 \E[ \psi(\fv,\Mm)] =  \sum_{f=1}^m \sum_{\ell=1}^U {U \choose \ell}  \rho_{f,\ell} \lambda(f,\ell),
\end{eqnarray}
where $\rho_{f,\ell}$ is given in \eqref{eq:rho}, from which the first term of the minimum in \eqref{eq:chi} follows. The second term follows directly from %, we reason similarly as in 
\cite{ji2015random,ji2015Efficient}.

If $\frac {\Delta}{\psi(\fv,\Mm) }  \geq   \frac { B D(1-p_{f_u} M ) }{\eta_{\mu(v)}}$, 
then $  \frac { B d_{\mu(v)} }{\Delta}=  \frac { B D(1-p_{f_u} M ) }{\Delta}
=\frac{(1-p_{f_u} M )}{\gamma}$, from which the third term in \eqref{eq:chi} follows, concluding the proof of  Theorem \ref{theorem: 1}.
%\textcolor{blue}{Please note that in the submitted version, there are 23 references, now there are 22 references. The reason is that the old reference [22] is not used,  (Y. H. K. A. El Gamal, Network Information Theory. )}
%-----------------------------------------------------------------------------------------------------------------------------------------------------------------------

\bibliographystyle{IEEEtran}
\bibliography{references_jsac}

\end{document}

%% file: macros.tex
\long\def\comment#1{}

\newfont{\bbb}{msbm10 scaled 700}

\newfont{\bb}{msbm10 scaled 1100}

\newcommand{\PP}{\mbox{\bb P}}

\newcommand{\FF}{\mbox{\bb F}}

\newcommand{\fv}{{\bf f}}

\newcommand{\pv}{{\bf p}}

\newcommand{\Mm}{{\bf M}}

\newcommand{\Pm}{{\bf P}}
\newcommand{\Qm}{{\bf Q}}

\newcommand{\Wm}{{\bf W}}

\newcommand{\Xm}{{\bf X}}

\newcommand{\Fc}{{\cal F}}

\newcommand{\Jc}{{\cal J}}
\newcommand{\Kc}{{\cal K}}

\newcommand{\Uc}{{\cal U}}

\newcommand{\Vc}{{\cal V}}

\newcommand{\fsf}{{\sf f}}

\newcommand{\Msf}{{\sf M}}

\renewcommand{\arg}{{\hbox{arg}}}

\newcommand{\eqdef}{\stackrel{\Delta}{=}}

\newcommand{\be}{\begin{equation}}
\newcommand{\ee}{\end{equation}}
\newcommand{\bea}{\begin{eqnarray}}
\newcommand{\eea}{\end{eqnarray}}

\def\E{{\mathbb E}}

\def\fsf{ {\sf f}}